\documentclass[preprint,showpacs,preprintnumbers,amsmath,amssymb]{revtex4}

\usepackage{graphicx,subfigure}
\usepackage{amsfonts}

\begin{document}

\title{Homology and symmetry breaking \\ in Rayleigh-B\'{e}nard convection:\\ Experiments and simulations}

\author{Kapilanjan Krishan}
 \altaffiliation[Present Address: ]{Department of Physics and Astronomy, University of California, Irvine, Irvine, CA 92697}
\author{Huseyin Kurtuldu}%
\author{Michael F. Schatz}%
\email{michael.schatz@physics.gatech.edu}
\affiliation{Center for Nonlinear Science and School of Physics, Georgia Institute of Technology, Atlanta, GA 30332}

\author{Marcio Gameiro$^{\ddag}$}
\author{Konstantin Mischaikow}
 \altaffiliation[Present Address:]{Rutgers University}
\affiliation{School of Mathematics, Georgia Institute of Technology, Atlanta, GA 30332}

\author{Santiago Madruga}
 \altaffiliation[Present Address:]{Max-Planck-Institute for Physics of Complex Systems, D-01187, Dresden, Germany}
\affiliation{Engineering Science and Applied Mathematics, Northwestern University, Evanston, IL 60208}

\date{\today}

\begin{abstract}

Algebraic topology (homology) is used to analyze the weakly turbulent state of spiral defect chaos 
in both laboratory experiments and numerical simulations of Rayleigh-B\'{e}nard convection.  The analysis reveals 
topological asymmetries that arise when non-Boussinesq effects are present.  The asymmetries are 
found in different flow fields in the simulations and are robust to substantial alterations to 
flow visualization conditions in the experiment.  However, the asymmetries are not
observable using conventional statistical measures.  These results suggest homology may provide
a new and general approach for connecting spatio-temporal observations of chaotic or turbulent patterns to  
theoretical models.

\end{abstract}
\maketitle

\section{Introduction}

Recent technical advances in experimental fluid mechanics 
now make it possible to measure complex dynamical behavior 
with high resolution in space and time (\cite{Bodenturb}).  
Similarly, modern 
computational fluid dynamics methods permit 
modelling of complex chaotic and turbulent flows (\cite{Kawahara}).  The data sets 
produced by such experiments and simulations can be enormous; as a result, interpreting 
the results becomes a significant challenge.  In particular, characterizing the geometric
properties of complex spatio-temporal patterns in large data sets has been
difficult because, to date, no general methodology has existed 
for extracting geometric signatures. 
    
Algebraic topology provides a tool for describing global 
geometric properties of
structures.  Devised by Poincar\'{e} \cite{Poincare} 
for use in global nonlinear analysis, 
algebraic topology originally used as input analytically defined 
objects ({\it{e.g.}}, level
sets of differentiable functions) to produce an output in the form of algebraic quantities
which convey topological information about the input.  In modern times, input objects can be expressed
either as simplicial or cubical complexes.  
In fluid mechanics and in most fields of science
and engineering, cubical representations often arise naturally in both experiments 
(raw image data
represented as square pixels or cubic voxels) and simulations (fields computed at gridpoints on square
or cubic lattices).  A package of computer programs has been developed to perform computations
of algebraic topology (computational homology) on cubical complexes in arbitrary 
dimensions.  This suite of tools, called {\tt{CHomP}} ({\tt{C}}omputational 
{\tt{Hom}}ology {\tt{P}}roject) is freely available for download via the Web (\cite{Hom,HomBook}).

We report here the first use of computational homology to characterize 
data obtained from a laboratory experiment.  We analyze Rayleigh-B\'{e}nard convection in 
the state known as spiral defect chaos (\cite{Morris93}), which is widely considered 
a paradigm for the little understood
phenomenon known as spatio-temporal chaos (\cite{Gollub}).  
In planform, patterns of spiral defect 
chaos, which are observed just above convective onset in
low Prandtl number ($\sim 1$) fluids, are composed of convection rolls deformed into numerous rotating
spirals and riddled with dislocations, disclinations and grain boundaries.  
Spiral defect chaos has been quantitatively described by a wide variety of
approaches, including structure factors, correlation lengths and times as well as wavenumber, spectral and spiral number distributions (\cite {madruga}; see also \cite {Bodenafrm} and references therein).

Thermal convection is frequently modelled using the Boussinesq approximation, which assumes the temperature dependence of the fluid properties can be neglected, except for the
temperature-induced density difference in the buoyant force that drives the flow.  However, non-Boussinesq effects
can arise in flows both in the laboratory and in natural settings.  At convective onset, the subcritical 
bifurcation to hexagonal patterns is a clear signature of 
non-Boussinesq effects. (Straight
convection rolls arise at onset from a supercritical bifurcation in Boussinesq Rayleigh-B\'{e}nard convection.)
Non-Boussinesq effects can be described quantitatively using perturbation theory near onset; in this
regime they are characterized by parameter $Q$ introduced by \cite{Busse}.  
Values of  $Q \geq \sim 1$ indicate significant non-Boussinesq effects; $Q = 0$ 
for Boussinesq convection.
As changes in control parameter move the convective flow well away from onset, 
non-Boussinesq effects typically
become more important and are more difficult to characterize 
theoretically and experimentally.

\section{Experimental procedure}\label{sec:experiment}

We measure convective flow in a horizontal layer of compressed (3.2 MPa absolute pressure) CO$_2$ gas 
of depth $d$ = 0.0690 $\pm$0.0005 cm.  The layer is bounded above by a 5-cm-thick sapphire window and below by 
1-cm-thick gold plated aluminum mirror. The lateral walls are circular, formed out of an
annular stack of filter paper sheets 3.80 cm in diameter. An electrical resistive heater
is used to heat the bottom mirror to a temperature, $T_h$, and the top window is cooled to a fixed temperature of $T_c$ =  21.20 $\pm$0.02~$^o$C by circulating chilled water. When the temperature difference, $\Delta T = T_h - T_c$, across the gas exceeds a critical temperature difference, $\Delta T_c$ = 3.8 $\pm$0.1~$^o$C, the onset of fluid motion occurs. The Prandtl number $Pr$ is 0.97.  In the experiments, the system control 
parameter, the reduced Rayleigh number 
$\epsilon = (\Delta T- \Delta T_c)/\Delta T_c$,
is increased above onset through a range where spiral defect chaos occurs.
The characteristic time scale, the vertical diffusion time $t_v$, is approximately 2 seconds.

\begin{figure}%
\begin{center}%
{\includegraphics[width=4.2cm]{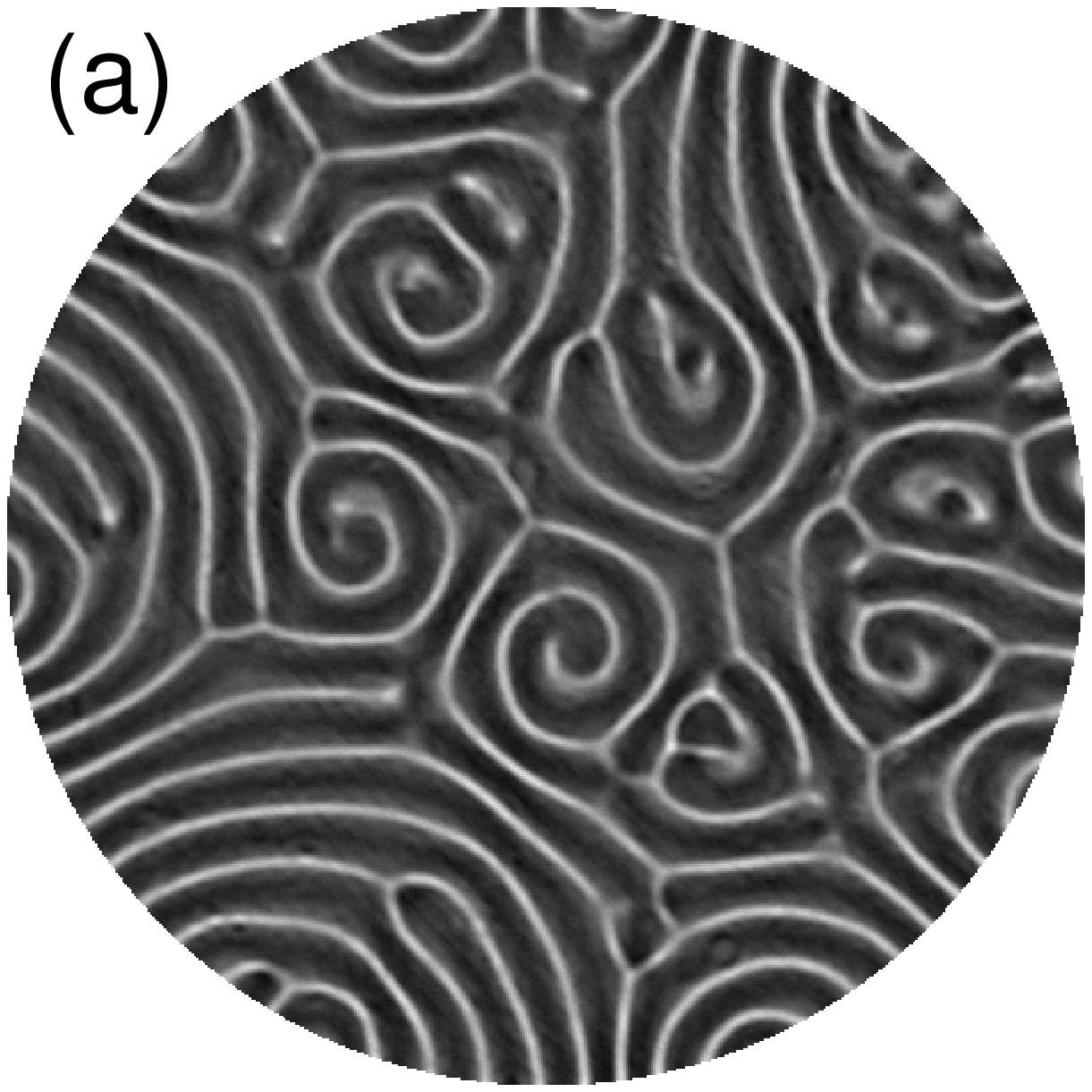}}
{\includegraphics[width=4.2cm]{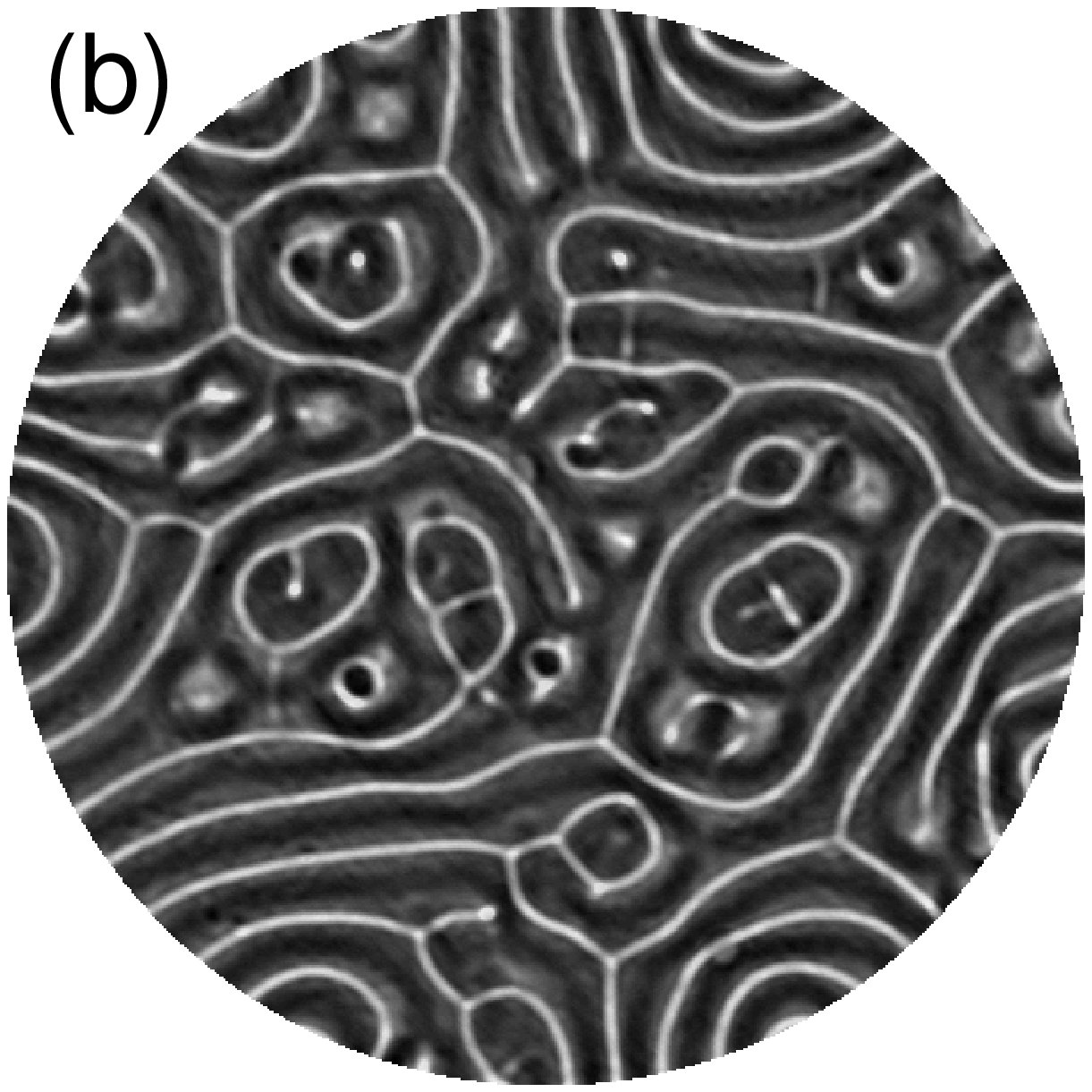}} \\
{\includegraphics[width=4.2cm]{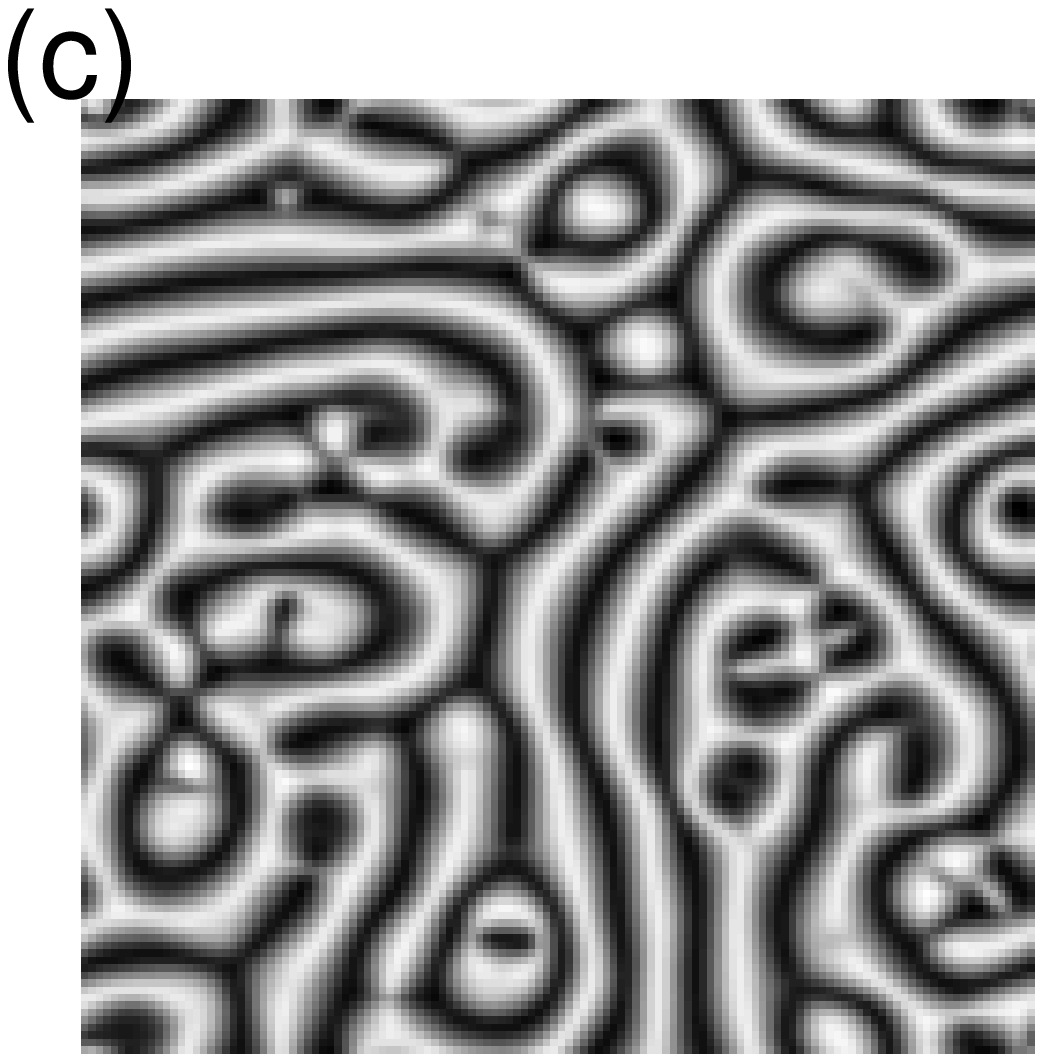}}
{\includegraphics[width=4.2cm]{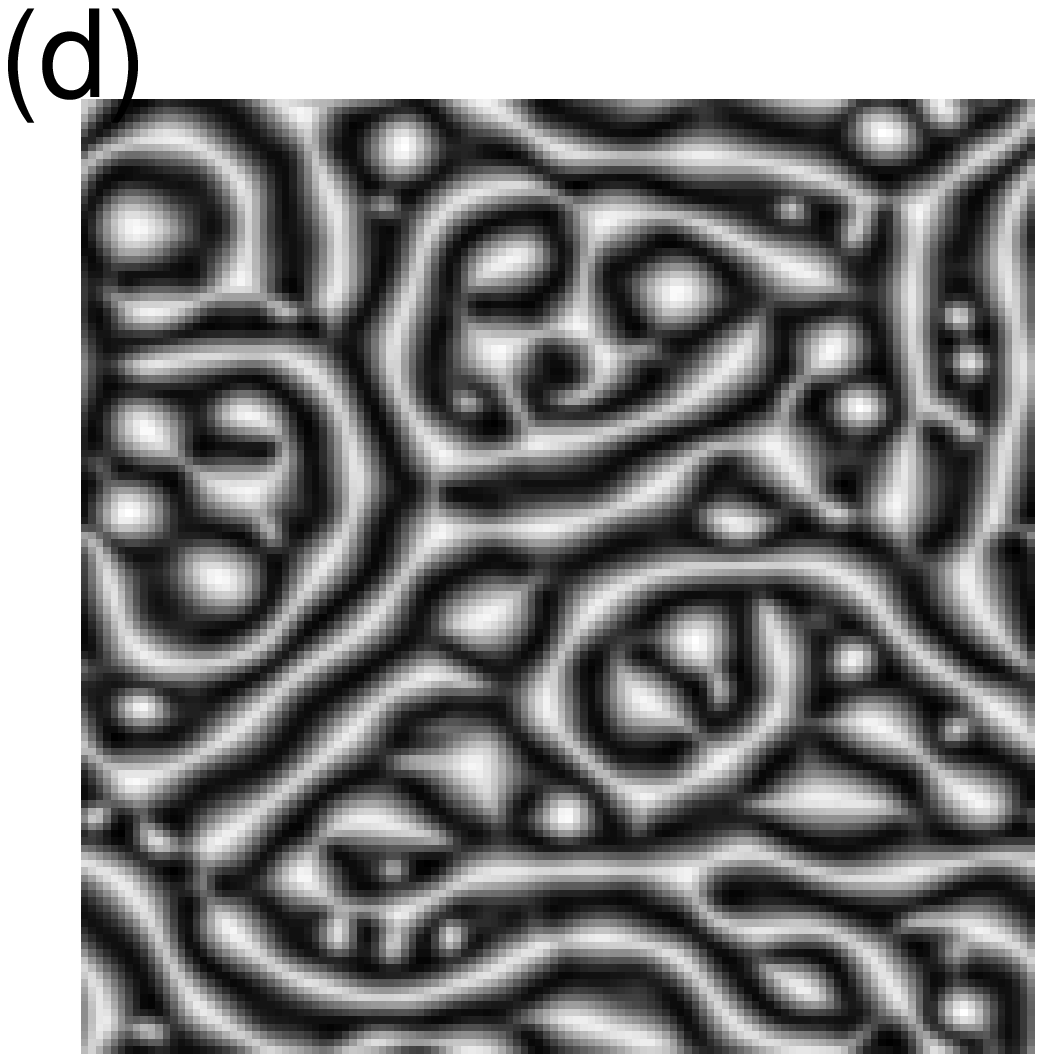}}
\caption{Images of spiral defect chaos convection are shown  
from laboratory experiments (a) \& (b)
and numerical simulations (c) \& (d).  Shadowgraph images from the experiments
illustrate the convective flows at (a) $\epsilon = 1.0$ and (b) $\epsilon = 2.0$.  The mid-plane temperature field is shown at $\epsilon =1.4$ for simulations
carried out under (c) Boussinesq ($Q=0$) and (d) Non-Boussinesq ($Q = 4.5$) conditions.  In all cases, bright regions in the images indicate the hot upflows
and dark regions indicate cold downflows in the convective patterns.}%
\label{sdc_images}%
\end{center}%
\end{figure}

The flows are visualized using the shadowgraph technique (\cite{debruyn}). Time series of shadowgraph images [Figure~\ref{sdc_images}\, ({a},{b})] 
with a spatial resolution of 515 x 650 pixels are captured under computer control 
at a rate of 11 Hz using a 12-bit digital camera interfaced to a framegrabber. The images are prepared for
analysis by subtracting a background image of the fluid below onset and using
digital Fourier filtering to remove high wavenumber components due to camera spatial
noise. The median value of intensity for all pixels in the image is then determined and used as a typical threshold to characterize each pixel as describing either hot upflow
or cold downflow in the convection pattern.  The resulting time series of 
thresholded 1-bit images are used for computing homology.

\section{Numerical simulations}\label{sec:numerics}
Our direct numerical simulations of the Navier-Stokes equations employ a pseudospectral code
developed by Pesch and co-workers (\cite{Peschprl,Peschchaos}).  
The code uses Fourier modes in the horizontal
direction and appropriate combinations of trigonometric and Chandrasekhar functions the satisfy the
top and bottom boundary conditions in the vertical direction \cite{Busse}.   
All runs are performed with
six vertical modes and 128 x 128 horizontal Fourier modes in a square domain with side length 
equal to 16 times the pattern wavelength at convective onset. The time step is typically $\tau_v$/500.  
For our analysis, the flows are
represented by 128 X 128 images [Figure~\ref{sdc_images}\, ({c},{d})] of the temperature field or the vertical velocity component.  
The images are typically stored every 2$\tau_v$. The median value of the flow field quantity (temperature or 
vertical velocity) for each image is 
determined and used as a threshold to characterize each gridpoint as describing hot upflow or cold downflow.  Thus, as in the lab experiment, the resulting 
time series of thresholded 1-bit images are used for computing homology.

In the simulations we describe non-Boussinesq effects arising from
the temperature dependence of material properties 
by a Taylor expansion truncated at leading-order 
beyond the Boussinesq approximation.  The simulations are performed 
at constant mean temperature $(T_h+T_c)/2$; the expansion is carried out 
about the mean temperature. 
In this case, the parameter $Q$ (\cite{Busse}) is given by

\begin{equation}
Q = \sum_{i=0}^{4}\gamma_i^{c} {\cal P}_i
\end{equation}

\noindent where the quantities ${\cal P}_i$ are linear functions of
$Pr^{-1}$, and the non-Boussinesq coefficients $\gamma_i^{c}$ give the
difference of the respective fluid properties across the layer at
threshold.  In non-Boussinesq simulations all the $\gamma_i^{c}$  
are retained, while in the Boussinesq simulations $\gamma_i^c$ are set to 0.
In all simulations, we fix $\epsilon = 1.4$ and set $Pr$ = 0.8.

\section{Results}\label{sec:results}

Formally, homology is computed for a topological space $X$ of N dimensions 
by assigning systematically a sequence of abelian groups $H_{k}(X)$ ($k=0,1,2,..., N-1$) 
to $X$.  For our purposes, it is sufficient 
to take $H_{k}$ to be products
of the integers: $H_{k}(X) = Z^{\beta_{k}(X)}$, where the integer dimensions of the groups $\beta_{k}(X) \geq 0$ 
are also known as the Betti numbers.  In this work we focus solely on $\beta_{k}(X)$ as the output
of the homological analysis; each $\beta_{k}(X)$ describes a topological property of $X$.
Thus, the net result of this analysis is the characterization of $X$ by a set of N non-negative integers.

\begin{figure}%
\begin{center}%
{\includegraphics[width=4.2cm]{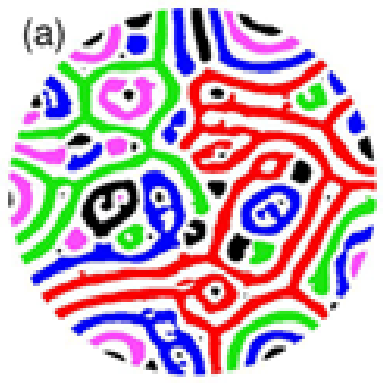}}
{\includegraphics[width=4.2cm]{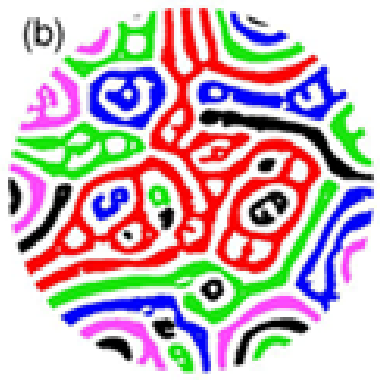}}\\
{\includegraphics[width=4.2cm]{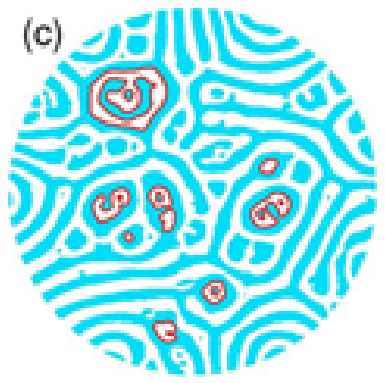}}
{\includegraphics[width=4.2cm]{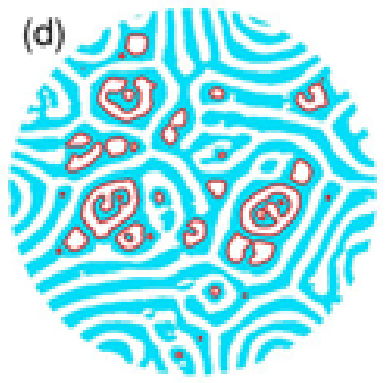}}
\caption{Computation of the homology for the
experimental data in Figure~\ref{sdc_images} ({b}) yields
a sequence of Betti numbers $\beta$, which can be readily interpreted
visually.  The number of distinct components is indicated by the 
zeroth Betti number for hot upflows (a) $\beta_{0h} = 65$ and for cold
downflows (b) $\beta_{0c} = 34$. (Different colors are used in (a) and (b)
 to distinguish a given component from its nearest neighbors.)
The number of holes is given by the first Betti number for
hot upflows (c) $\beta_{1h} = 9$ and for cold downflows 
(d) $\beta_{1c} = 34$. (Each hole is encircled by a red boundary in (c) and (d).)}%
\label{hom_patterns}%
\end{center}%
\end{figure}

From each 1-bit image in time series from either experiments or simulations, two distinct, topological spaces 
are obtained: $X_{h}$, where the hot upflow pixels have non-zero value and $X_{c}$ where the cold 
downflow pixels have non-zero value.  $X_c$ and $X_h$, which are two-dimensional, are input, in turn, 
into the homology codes, which subsequently output two Betti numbers for each space: $\beta_{0h}$, 
$\beta_{1h}$ for the hot upflows and $\beta_{0c}$, $\beta_{1c}$ for the cold downflows.  
$\beta_{0h}$ ($\beta_{0c}$) counts the number of distinct components, i.e., the number of regions of hot 
upflow (cold downflow) that are separated from similar regions in a given pattern $X_h$ ($X_c$) [Figure~\ref{hom_patterns}\, ({a},{b})].  
$\beta_{1h}$ ($\beta_{1c}$) counts the number of holes in the hot 
upflows (cold downflows) in a given pattern $X_h$ ($X_c$) [Figure~\ref{hom_patterns}\, ({c},{d})].  
With the package {\tt CHomP}, computing the homology of $X_c$ and $X_h$ corresponding
to each image takes about 1 second.

Figure~\ref{hom_patterns} shows a striking result: in the experiments,
hot upflows are topologically quite distinct from cold downflows.
Specifically, there are more hot upflow components than cold downflow components
($\beta_{0h} > \beta_{0c}$).  Moreover, the cold downflow regions contain more holes
than the hot upflows ($\beta_{1c} > \beta_{1h}$).  This distinction is not revealed using 
standard statistical measures of the pattern.  For example, the mean area occupied by upflow
is equal to that occupied by downflow by construction (when the threshold is set to
the median pixel intensity in the original image.)  Wavenumber distributions obtained from
Fourier analysis of $X_h$ and $X_c$ show no discernible differences.

These measurements of topology are robust to variations in the choice of 
threshold.    The choice of the median pixel intensity
as the threshold to separate upflows from downflows is physically well-motivated but somewhat arbitrary.  In practice,
any reasonable choice yields similar results.  For example, for $X_h$ and $X_c$ in Figure 2, choosing the mean pixel intensity (which is larger than the median intensity by approximately 5\% of full scale) as the threshold yields nearly 
identical Betti numbers:  $\beta_{0h} = 65$, $\beta_{0c} = 34$,
$\beta_{1h} = 9$, $\beta_{1c} = 33$.

\begin{figure}%
\begin{center}%
{\includegraphics[width=6.5cm,height=3.5cm]{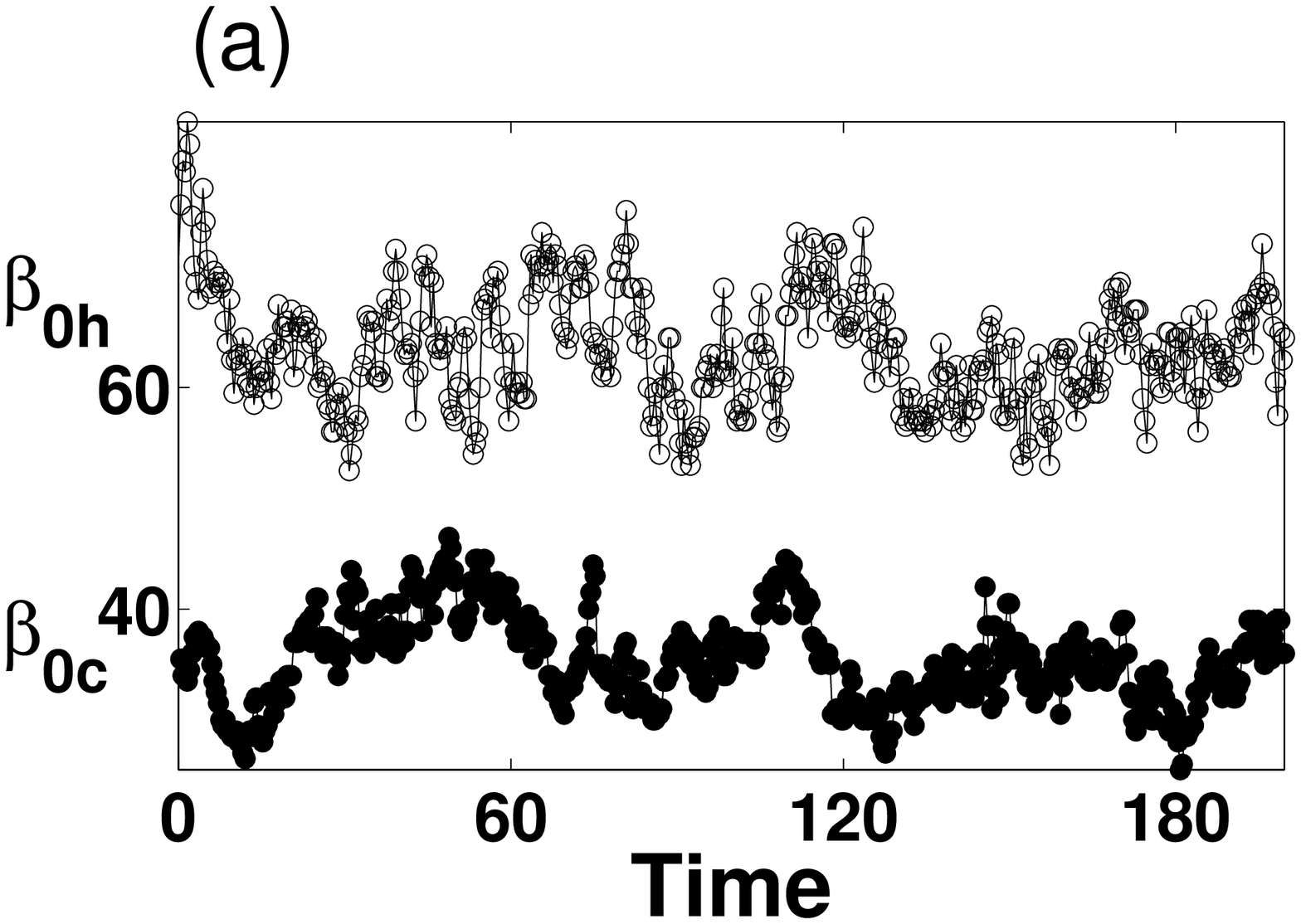}}
{\includegraphics[width=6.5cm,height=3.5cm]{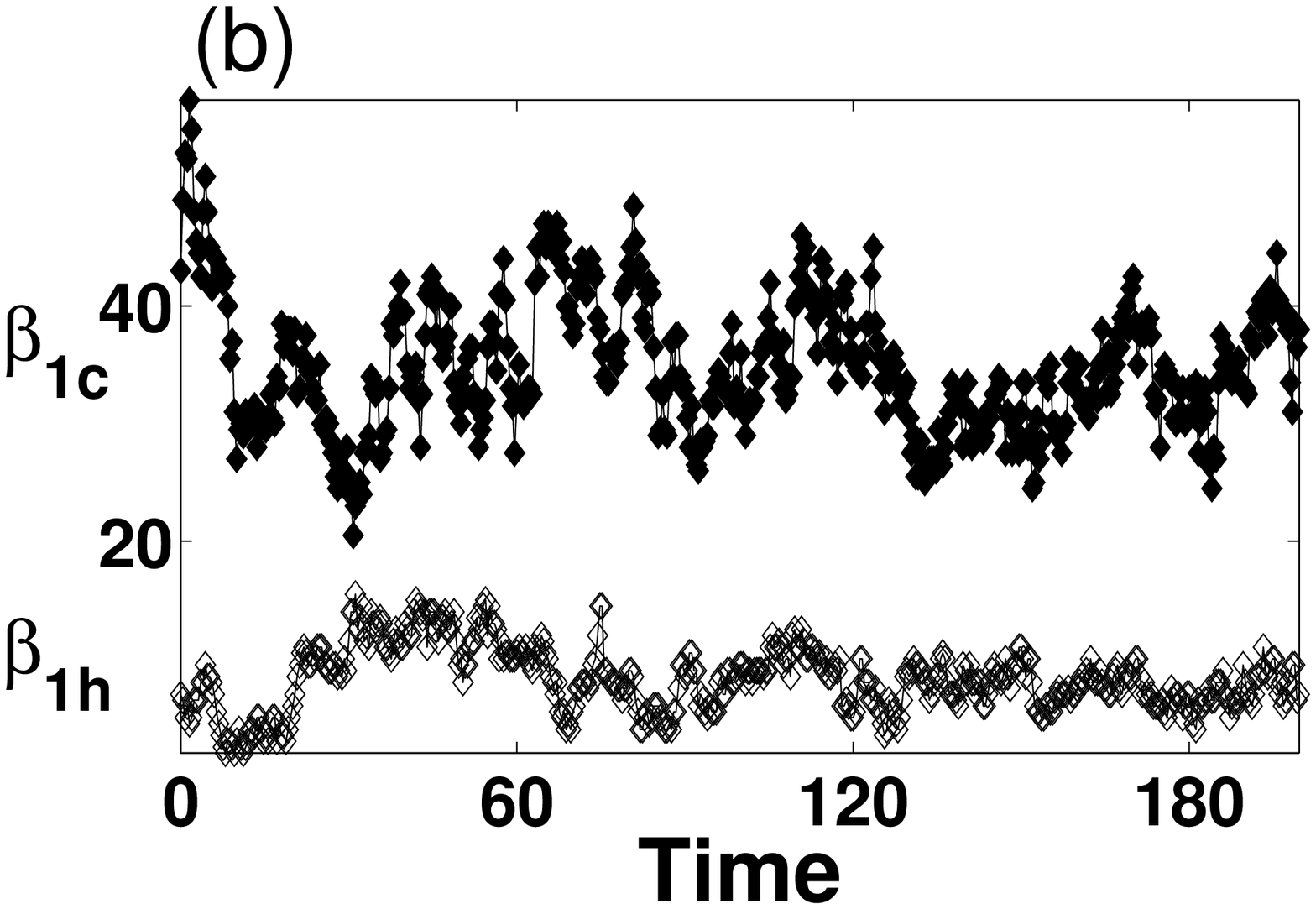}}
\caption{Time series of (a) the zeroth Betti numbers $\beta_{0h}$ (open circles), $\beta_{0c}$ (filled circles) and (b) the first Betti numbers $\beta_{1h}$ (open diamonds) and $\beta_{1c}$ (filled diamonds) are obtained from laboratory experiments at $\epsilon = 2.0$. Time is scaled by $t_v$; the time interval between samples is $t_v/2$}%
\label{betai_es}%
\end{center}%
\end{figure}

Time series of the Betti numbers exhibit fluctuations about 
well-defined time average values (Figure~\ref{betai_es}).  
The fluctuations are primarily a global signature 
of the complex spatiotemporal behavior of spiral defect chaos.  Mean flow induced by 
curvature in the
roll pattern leads to regions of
local compression or dilatation throughout the pattern.  Compression often leads to mergering
of neighboring rolls while the dilatation results in the formation of a new rolls in the
pattern; these processes are closely related to secondary instability mechanisms for ideal straight 
rolls [\cite{Busse,Bodenafrm}].  These local events drive further changes in pattern curvature, thereby
leading to a continually evolving pattern with fluctuating topology.  The Betti numbers are a global
measure of the topological changes, and therefore, are dependent on the local processes, for which
theories of defect dynamics have been proposed [\cite{crosshohen}].  How Betti numbers are related to defect dynamics
remains an open question; for our purposes here, we focus on the time average values of the Betti numbers ($\bar\beta_{0h}$,$\bar\beta_{1h}$,$\bar\beta_{0c}$,$\bar\beta_{1c}$), which
we find to be stationary for fixed $\epsilon$.

\begin{figure}%
\begin{center}%
{\includegraphics[width=8cm]{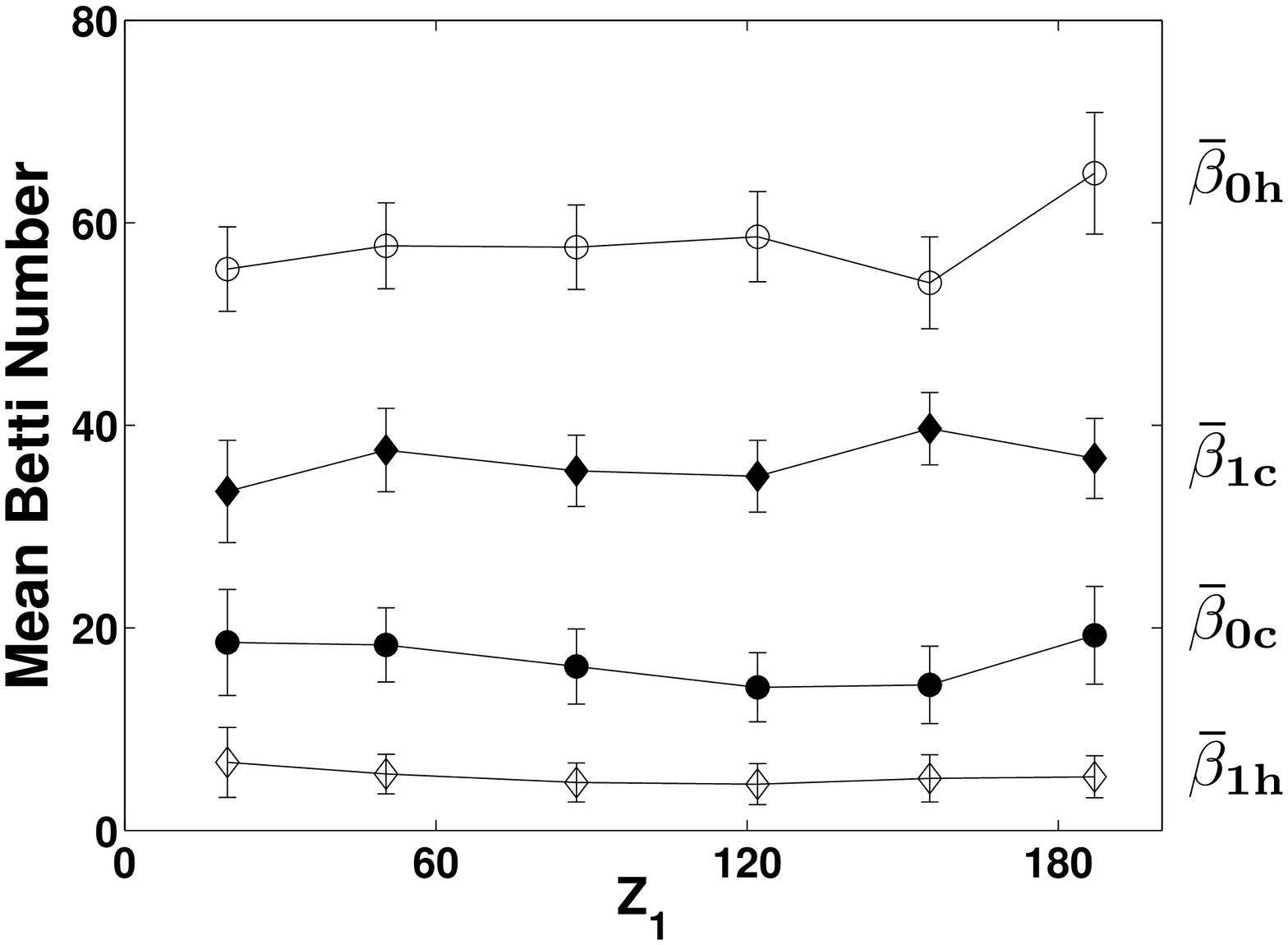}}
\caption{The mean Betti numbers are plotted as a function of the effective 
optical distance $z_1$ of the shadowgraph system in laboratory experiments
performed at $\epsilon = 2$. For each data point, the median pixel
intensity of the raw shadowgraph images was used as the threshold for
the homology analysis.}%
\label{betti_vs_z1}%
\end{center}%
\end{figure}

The measurements of $\bar\beta$ are robust with respect to flow visualization conditions.  
It is well-known that shadowgraphy can introduce significant nonlinearities and 
image artifacts (e.g., caustics); the strength of these effects depend on the effective optical distance $z_1$ of the shadowgraph system [\cite {debruyn}].  We have checked for possible sensitivity to shadowgraphy visualization 
by conducting a series of experiments where the conditions of the convective flow were fixed and 
image time series were captured for for different values of  $z_1$.   
Figure~\ref{betti_vs_z1} shows that 
the mean Betti number change only slightly as $z_1$ is varied over nearly an order of magnitude.  Additional experimental data (not shown) demonstrate that
a change of sign in $z_1$ (which changes hot upflows (cold downflows) from bright (dark) to dark (bright))
does not affect the determination of $\bar\beta$.

\begin{figure}%
\begin{center}%
{\includegraphics[width=8cm]{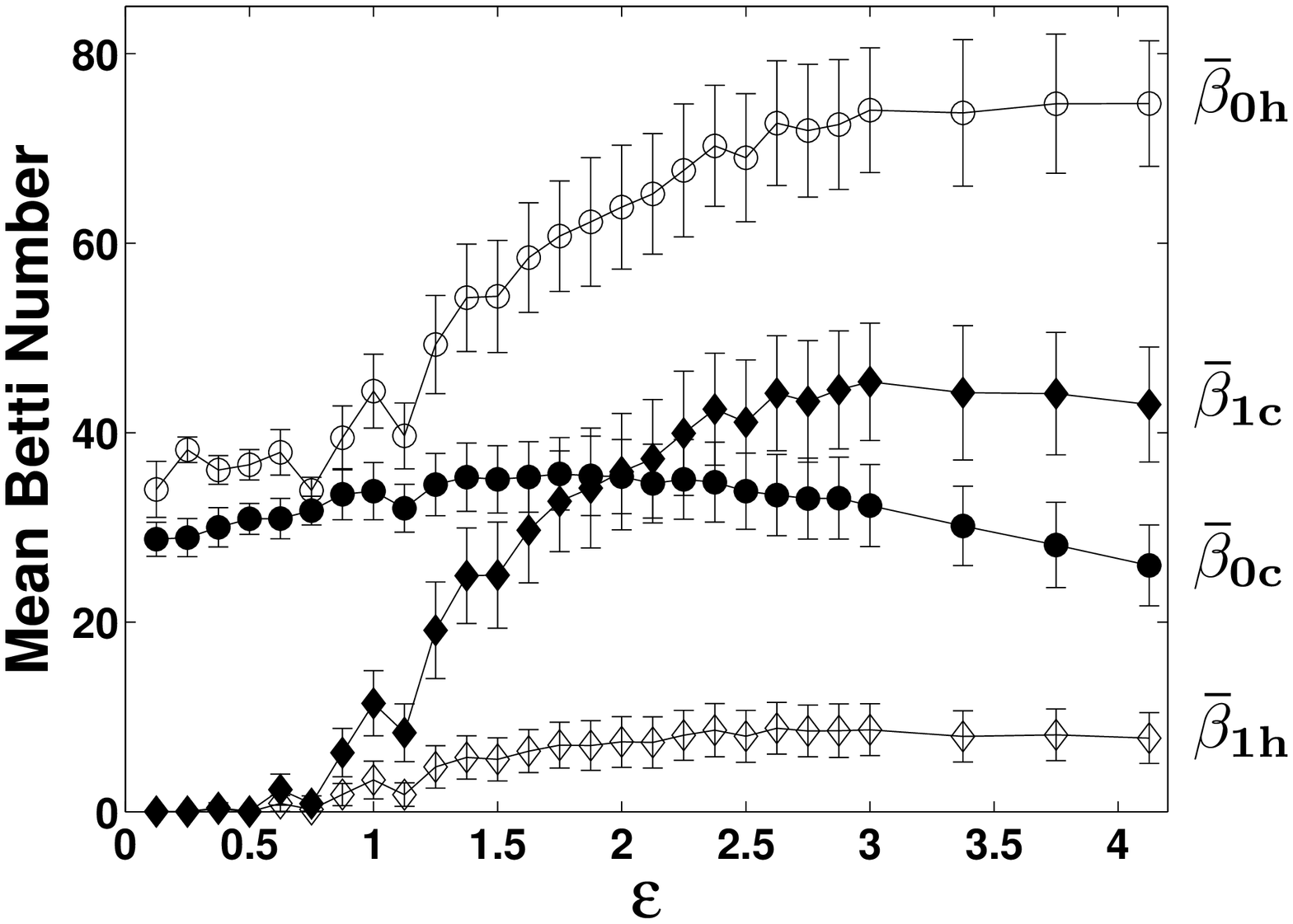}}
\caption{The mean zeroth Betti numbers $\bar\beta_{0h}$ (open circles), $\bar\beta_{0c}$ (filled circles)  and first Betti numbers $\bar\beta_{1h}$ (open diamonds) and $\bar\beta_{1c}$ (filled diamonds) are shown as a function of $\epsilon$ for data from laboratory experiments. Each data point is obtained by averaging the Betti numbers from analysis of 18000 images corresponding to an observation time of approximately 1800 $t_v$}%
\label{mean_values}%
\end{center}%
\end{figure}

The differences between the mean Betti numbers for hot upflows $\bar\beta_{0h},\bar\beta_{1h}$ and for cold down flows
$\bar\beta_{0c},\bar\beta_{1c}$ become more substantial as $\epsilon$ increased above convective onset (Figure~\ref{mean_values}).
For $\epsilon < 0.7$, the number of upflow components exceeds the 
number of downflow components by a small but statistically
significant amount while the number of holes in both types of flow patterns are effectively zero.  (The onset of spiral defect chaos occurs approximately
at $\epsilon = 0.7$ in our experiment.)  For $\epsilon > 0.7$, the difference in the average component number grows significantly; at first 
primarily because the number of hot components grows rapidly for $0.7 < \epsilon < 2.5$ and then because
the number of cold components decreases for $2.5 < \epsilon < 4$.  The behavior in the number of holes
is somewhat different.  In the range $0.7 < \epsilon < 2.5$ the number of holes increases significantly 
in the cold downflows, but only weakly for the hot upflows.  
  
\begin{figure}%
\begin{center}%
{\includegraphics[width=6.5cm,height=3.5cm]{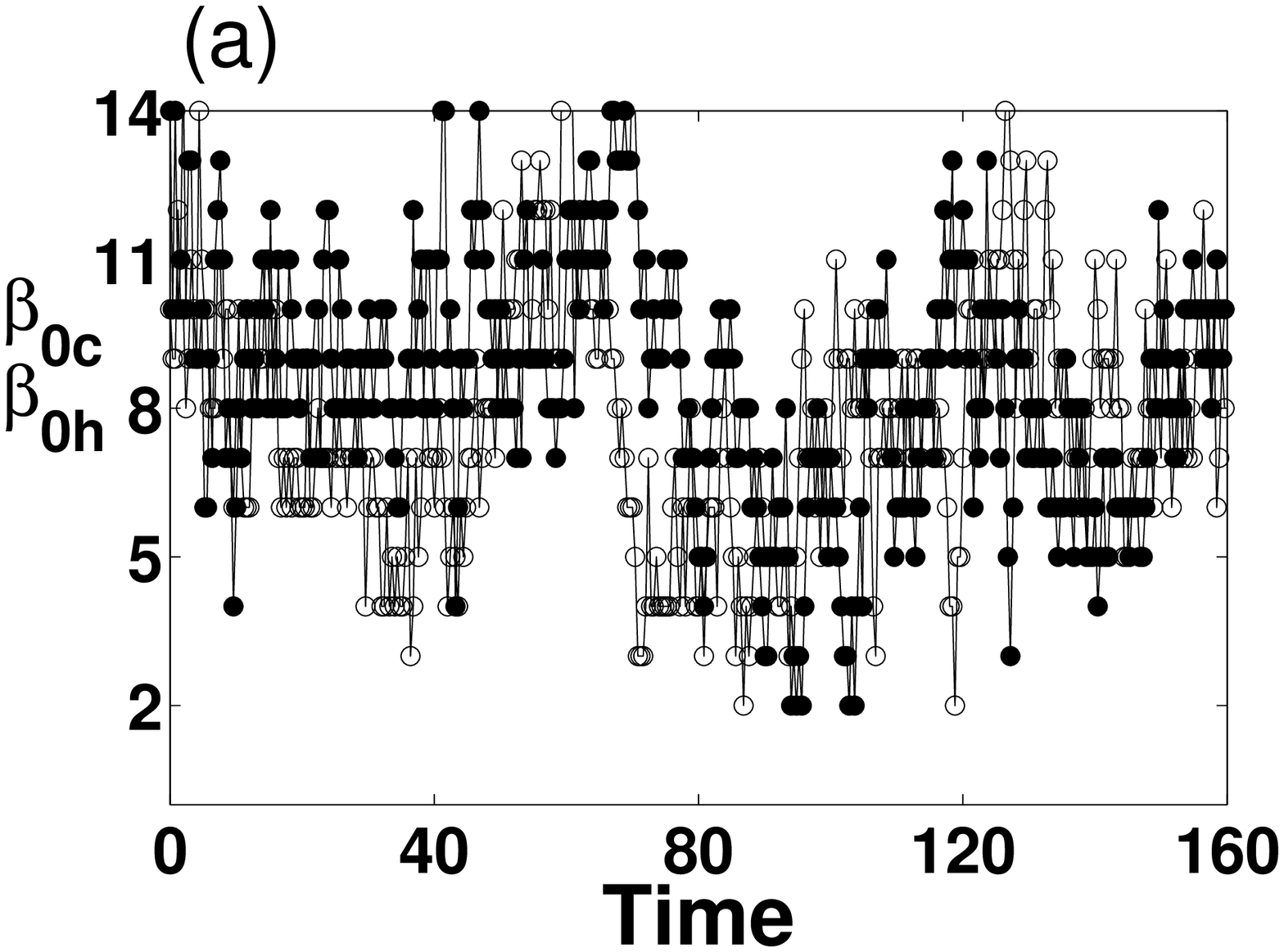}}
{\includegraphics[width=6.5cm,height=3.5cm]{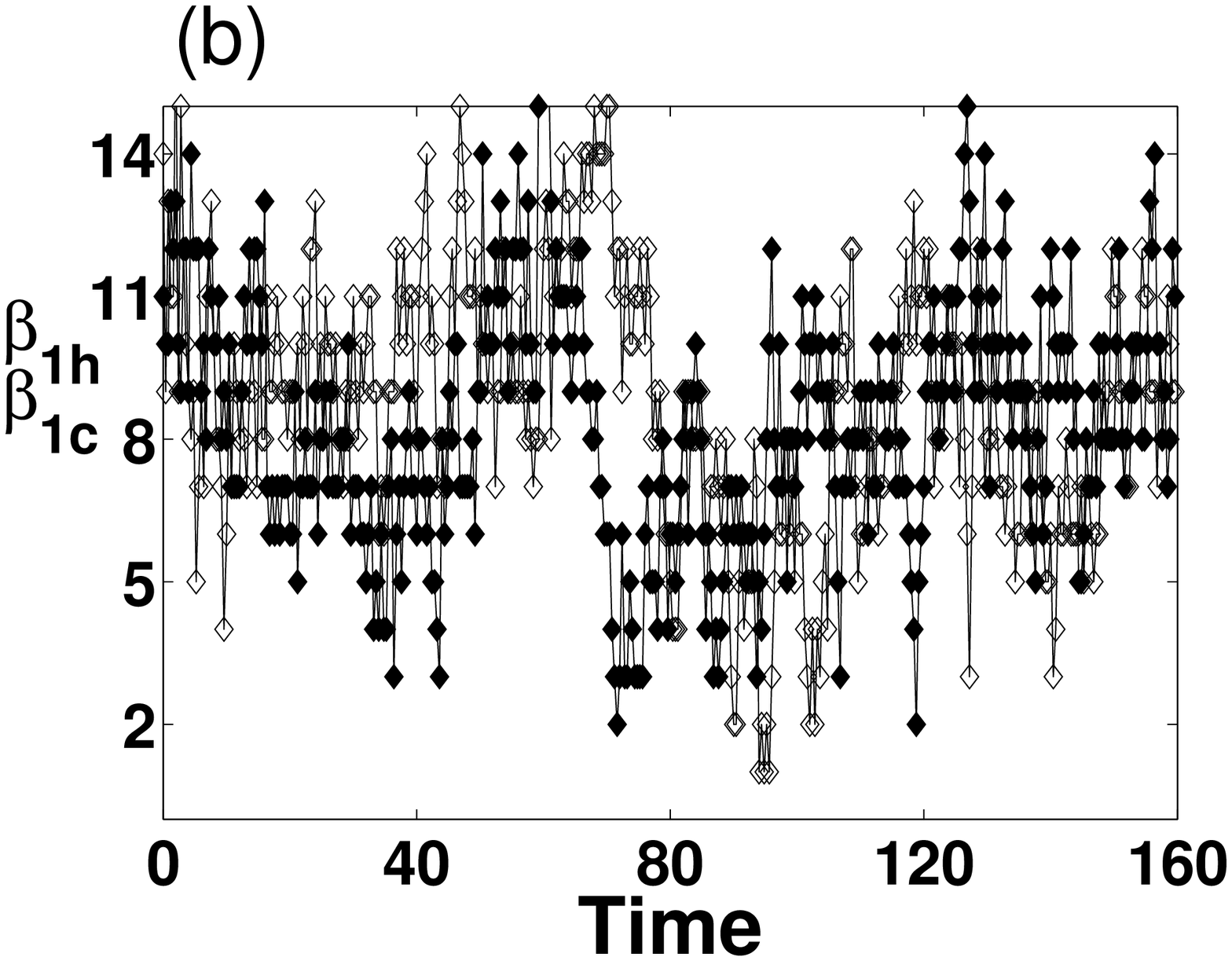}} \\
{\includegraphics[width=6.5cm,height=3.5cm]{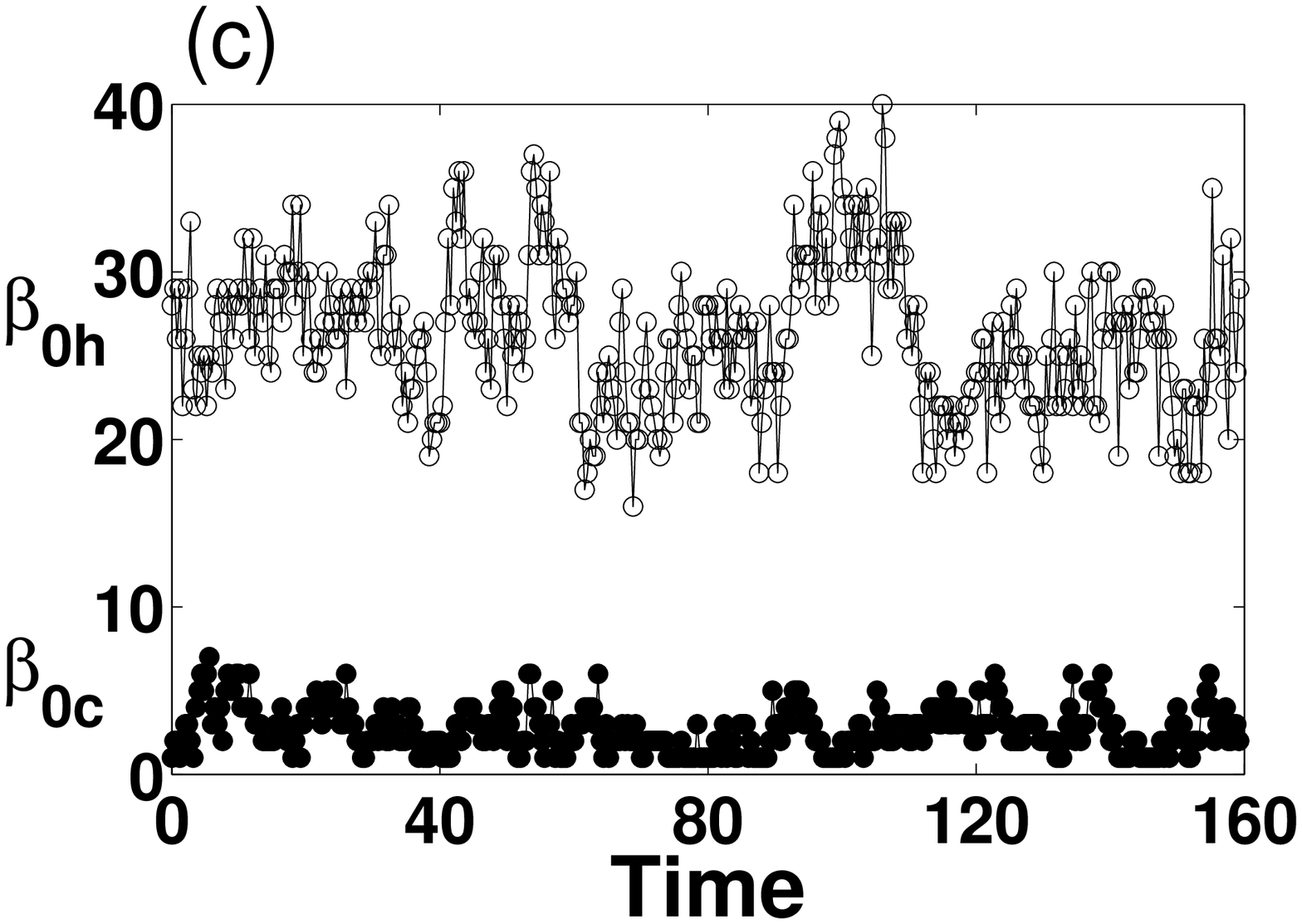}}
{\includegraphics[width=6.5cm,height=3.5cm]{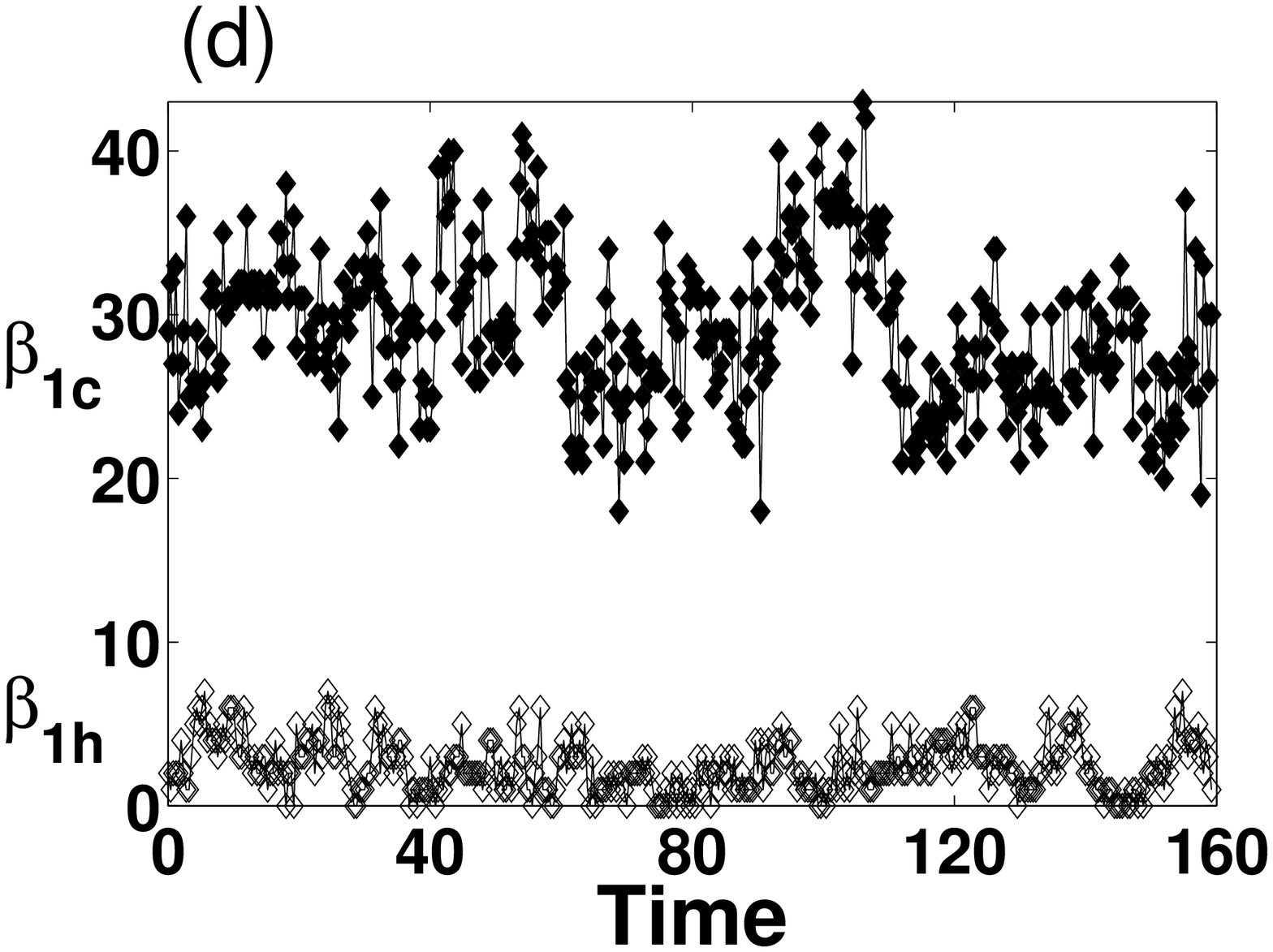}} \\
{\includegraphics[width=6.5cm,height=3.5cm]{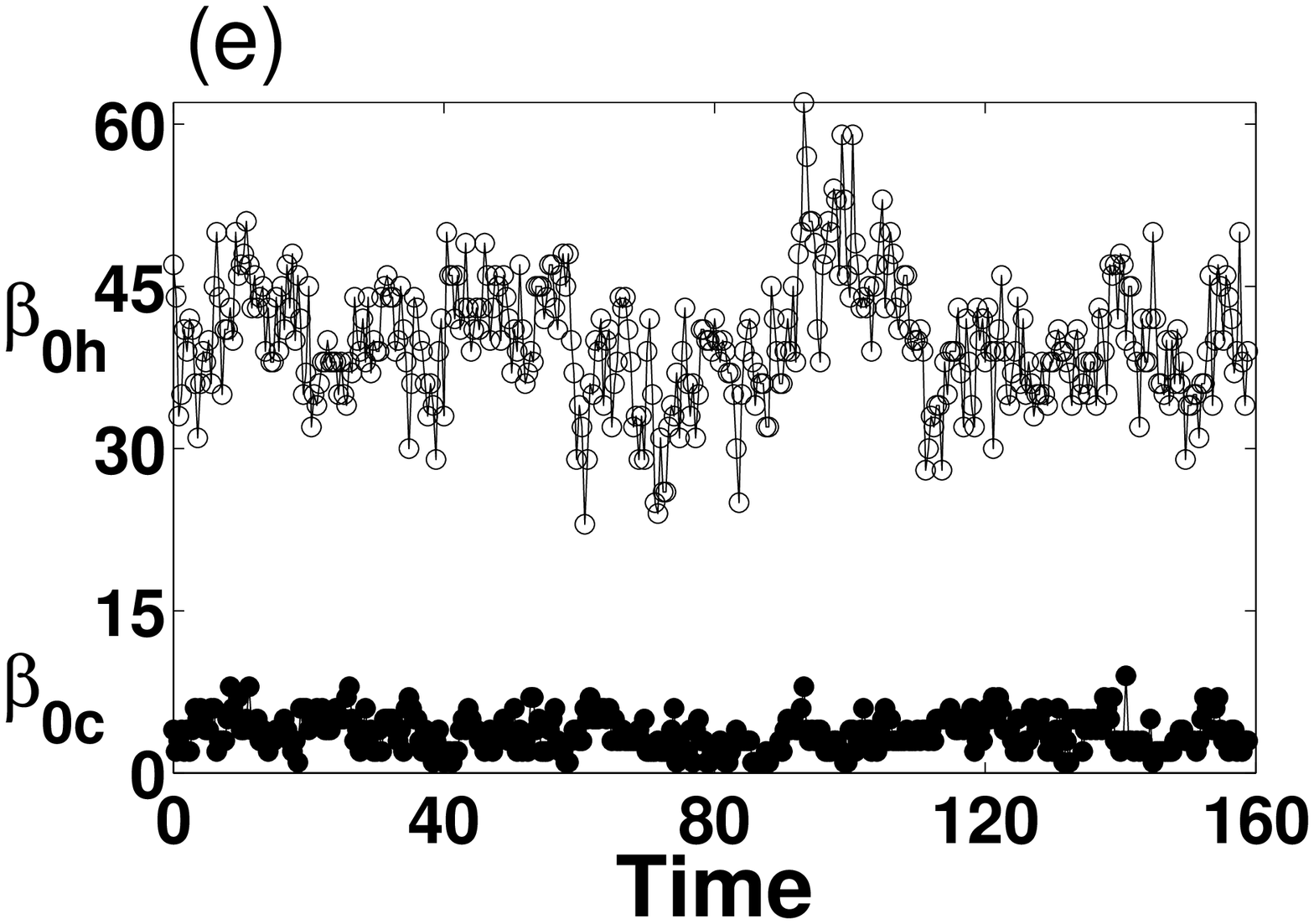}}
{\includegraphics[width=6.5cm,height=3.5cm]{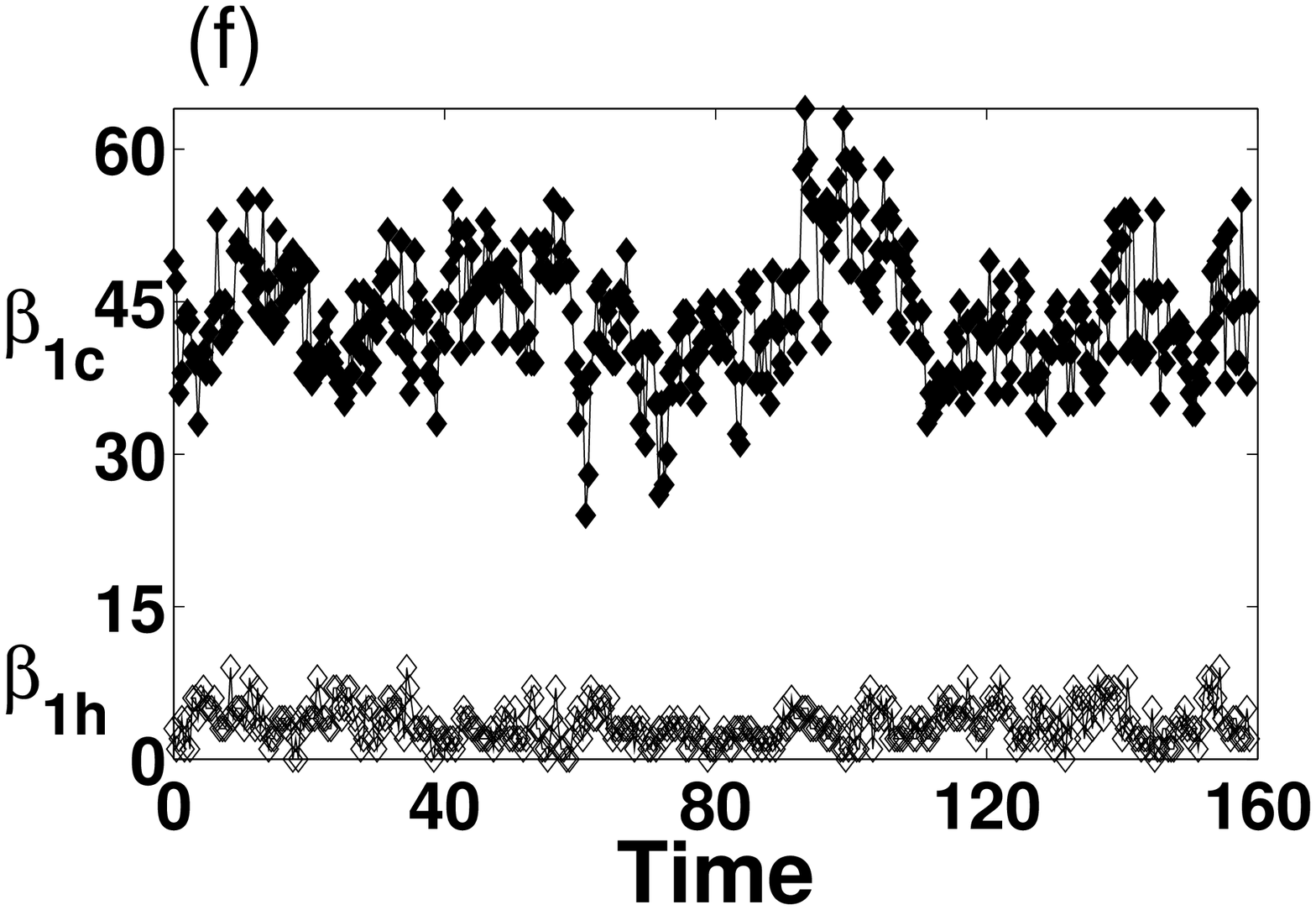}}
\caption{Time series of the zeroth Betti numbers  $\beta_{0h}$ (open circles), $\beta_{0c}$ (filled circles) and the first Betti numbers $\beta_{1h}$ (open diamonds) and $\beta_{1c}$ (filled diamonds) are obtained from numerical simulations at $\epsilon = 1.4$.  The midplane temperature field from Boussinesq simulations is used to obtain time series of (a) $\beta_{0h}$, $\beta_{0c}$ 
and (b) $\beta_{1h}$, $\beta_{1c}$. The midplane temperature field from non-Boussinesq simulations is used to obtain time series of (c) $\beta_{0h}$, $\beta_{0c}$ and (d) $\beta_{1h}$, $\beta_{1c}$. The vertical velocity component at
$z=-0.25$ from non-Boussinesq simulations is used to obtain time series of (e) $\beta_{0h}$, $\beta_{0c}$ and (f) $\beta_{1h}$, $\beta_{1c}$.  (The midplane
is located at $z=0$ and the bottom boundary is located at $z=-0.5$.)  Time
is scaled by $t_v$.}%
\label{betai_sim}%
\end{center}%
\end{figure}

These experimental observations, taken as a whole, suggest the observed asymmetries in the Betti numbers 
may be due to the breakdown of the Boussinesq approximation (and corresponding breaking of the 
Boussinesq symmetry).  To check this hypothesis, we conducted two simulations under identical conditions
except one simulation was Boussinesq ($Q=0$) and the other simulation was non-Boussinesq ($Q=4.5$).  
The Betti number time series for the Boussinesq simulation shows little distinction in the
Betti numbers [Figure~\ref{betai_sim}\, (a, b)].   By contrast, examination of
the same field variable in non-Boussinesq simulation [Figure~\ref{betai_sim}\, (c, d)]  shows distinct differences in the Betti numbers that are qualitatively in agreement with the experimental observations.

Different flow fields extracted from the simulation exhibit similar qualitative behavior in 
the time-averaged Betti numbers [Figure~\ref{betai_sim}\, (c, d, e, f)].  
We examined both the temperature field
and the vertical velocity components sampled at the vertical positions $z=0$ 
(the mid-plane of the convection cell), $z=-0.25$ and
$z=+0.25$.  (The top and bottom boundaries are located at $z=+0.5$ and $z=-0.5$, respectively.)
The quantitative values for the time-averaged Betti numbers differ weakly between different projections of the 
convective flow.  However,  every projections exhibit the same
qualitative result, namely, in a given projection of Boussinesq convection, the time averaged Betti numbers 
for hot upflows and cold downflows are the same whilst each projection of non-Boussinesq exhibits the same Betti number asymmetries.

\section{Conclusions}\label{sec:concl}

We conclude that the breakdown of the Boussinesq approximation can be
readily observed in data from convection experiments and simulations by 
analyzing the topology using computational homology.  It might be argued the use
of homology constitutes an "excessive use of force" for the two-dimensional
patterns analyzed here since the counting of features such as components and holes could be accomplished by
other means.  Nevertheless, the simple fact remains that upflow/downflow asymmetries, which had remained
unnoticed despite decades of study of convective flows, were uncovered 
precisely because of the solid mathematical foundation that
the homology formalism provides.  Moreover, the homology analysis outlined here can be readily extended 
to higher dimensions where less sophisticated approaches will likely fail.  For example, 
three-dimensional complexes can be formed from the image data used here by creating "time-blocks" of
data with two spatial dimensions and one time dimension; such data is expected to contain new topological
features that capture dynamical information (\cite{Gameiro}).

Describing the physical mechanism that connects the observed topological asymmetries to the 
breakdown of the Boussinesq approximation remains an open theoretical question.  There exist
physical mechanisms that connect the details of non-Boussinesq effects to the hexagonal flow states 
observed near onset; in particular, the hexagons observed in gases are "down-hexagons", i.e., they
have cold downflow in the center of each hexagon.  The physical mechanism behind this pattern selection
is attributed primarily to the fact that as the temperature increases in gases, the kinematic viscosity 
typically decreases and the thermal expansion coefficient usually increases [\cite{Busse}].  The zeroth Betti 
numbers
of a pattern of $M$ ordered hexagons can be easily determined.  The centers of each hexagon will be
isolated for all other cold downflows, yielding $\beta_{0c} =M$, while the hot upflows around the edges
of all hexagons will be connected, yielding $\beta_{0h} = 1$.  This behavior of the Betti numbers
$\beta_{0c} > \beta_{0h}$ is distinctly different from the results from experiments and
simulations at higher $\epsilon$, namely $\beta_{0h} > \beta_{0c}$.  A convincing physical explanation must
account for these differences.

Our results suggest that computational homology might be a useful tool in a wide variety of 
cases in fluid dynamics.  For example, in the atmospheric sciences, where extensive use is made of 
the Boussinesq approximation, homological analysis may provide new ways to characterize atmospheric data.
The use of homology need not be limited to convection; this approach may be applied
in any fluid flow where quantitative characterization of complex data is needed.



%

\begin{acknowledgments}
We would like to thank G. Ahlers, G. Gunaratne, B. Kalies, and H. Riecke 
for their helpful comments about this work.   This work was supported 
by DARPA (KM and MG), the Department of Energy under Grant 97891, and
the National Science Foundation under grants DMS-0107396 (KM and MG), CTS-0201610 (MFS and KK), DMS-0443827 (KM), ATM-0434193 (MFS and HK), and DMS-0511115 (KM).  MG also acknowledges the support of a CAPES (Brazil) fellowship.  
\end{acknowledgments}


\vfill\eject

\noindent{\bf FIGURE CAPTIONS}


{\bf Figure 1.} Images of spiral defect chaos convection are shown  
from laboratory experiments (a) \& (b)
and numerical simulations (c) \& (d).  Shadowgraph images from the experiments
illustrate the convective flows at (a) $\epsilon = 1.0$ and (b) $\epsilon = 2.0$.  The mid-plane temperature field is shown at $\epsilon =1.4$ for simulations
carried out under (c) Boussinesq ($Q=0$) and (d) Non-Boussinesq ($Q = 4.5$) conditions.  In all cases, bright regions in the images indicate the hot upflows
and dark regions indicate cold downflows in the convective patterns.

\medskip
\medskip

{\bf Figure 2.} Computation of the homology for the
experimental data in Figure~\ref{sdc_images} ({b}) yields
a sequence of Betti numbers $\beta$, which can be readily interpreted
visually.  The number of distinct components is indicated by the 
zeroth Betti number for hot upflows (a) $\beta_{0h} = 65$ and for cold
downflows (b) $\beta_{0c} = 34$. (Different colors are used in (a) and (b)
 to distinguish a given component from its nearest neighbors.)
The number of holes is given by the first Betti number for
hot upflows (c) $\beta_{1h} = 9$ and for cold downflows 
(d) $\beta_{1c} = 34$. (Each hole is encircled by a red boundary in (c) and (d).)

\medskip
\medskip


{\bf Figure 3.} Time series of (a) the zeroth Betti numbers $\beta_{0h}$ (open circles), $\beta_{0c}$ (filled circles) and (b) the first Betti numbers $\beta_{1h}$ (open diamonds) and $\beta_{1c}$ (filled diamonds) are obtained from laboratory experiments at $\epsilon = 2.0$. Time is scaled by $t_v$; the time interval between samples is $t_v/2$

\medskip
\medskip


{\bf Figure 4.} The mean Betti numbers are plotted as a function of the effective 
optical distance $z_1$ of the shadowgraph system in laboratory experiments
performed at $\epsilon = 2$. For each data point, the median pixel
intensity of the raw shadowgraph images was used as the threshold for
the homology analysis.

\medskip
\medskip

\vfill\eject

\noindent{\bf FIGURE CAPTIONS (cont.)}


{\bf Figure 5.} The mean zeroth Betti numbers $\bar\beta_{0h}$ (open circles), $\bar\beta_{0c}$ (filled circles)  and first Betti numbers $\bar\beta_{1h}$ (open diamonds) and $\bar\beta_{1c}$ (filled diamonds) are shown as a function of $\epsilon$ for data from laboratory experiments. Each data point is obtained by averaging the Betti numbers from analysis of 18000 images corresponding to an observation time of approximately 1800 $t_v$.


{\bf Figure 6.} Time series of the zeroth Betti numbers  $\beta_{0h}$ (open circles), $\beta_{0c}$ (filled circles) and the first Betti numbers $\beta_{1h}$ (open diamonds) and $\beta_{1c}$ (filled diamonds) are obtained from numerical simulations at $\epsilon = 1.4$.  The midplane temperature field from Boussinesq simulations is used to obtain time series of (a) $\beta_{0h}$, $\beta_{0c}$ and (b) $\beta_{1h}$, $\beta_{1c}$. The midplane temperature field from non-Boussinesq simulations is used to obtain time series of (c) $\beta_{0h}$, $\beta_{0c}$ and (d) $\beta_{1h}$, $\beta_{1c}$. The vertical velocity component at
$z=-0.25$ from non-Boussinesq simulations is used to obtain time series of (e) $\beta_{0h}$, $\beta_{0c}$ and (f) $\beta_{1h}$, $\beta_{1c}$.  (The midplane
is located at $z=0$ and the bottom boundary is located at $z=-0.5$.)  Time
is scaled by $t_v$.

\vfill\eject

\begin{figure}%
\begin{center}%
{\includegraphics[width=4.2cm]{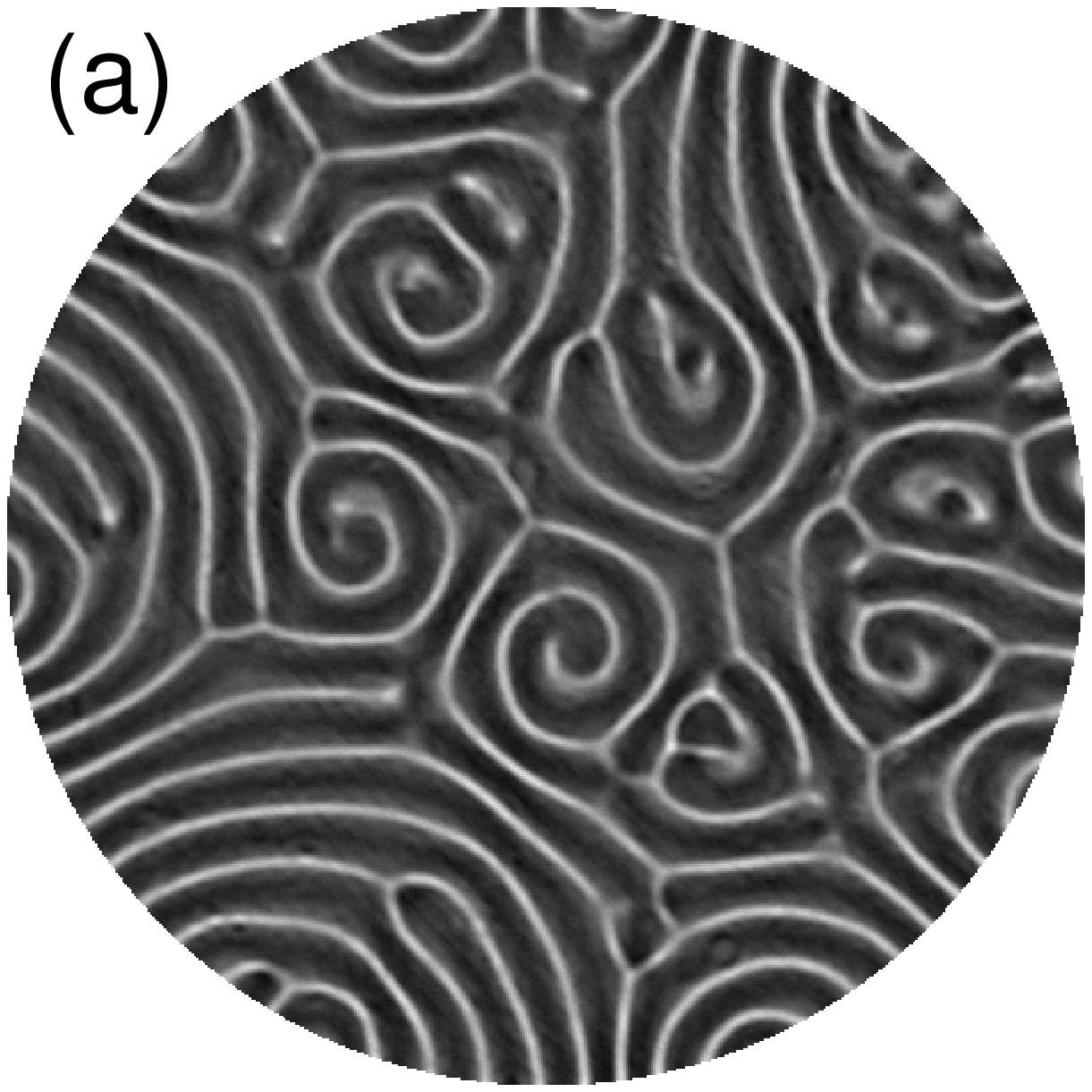}}
{\includegraphics[width=4.2cm]{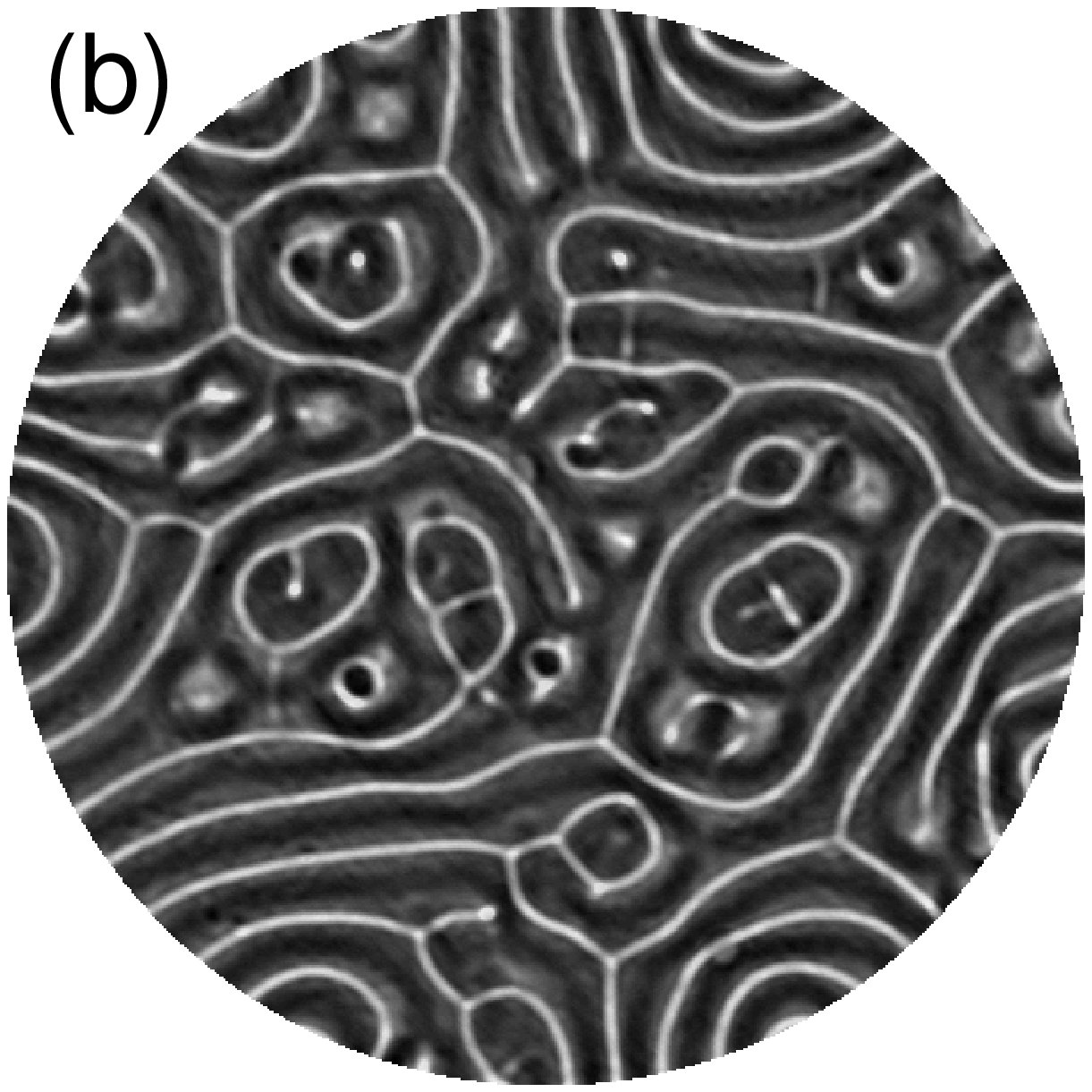}} \\
{\includegraphics[width=4.2cm]{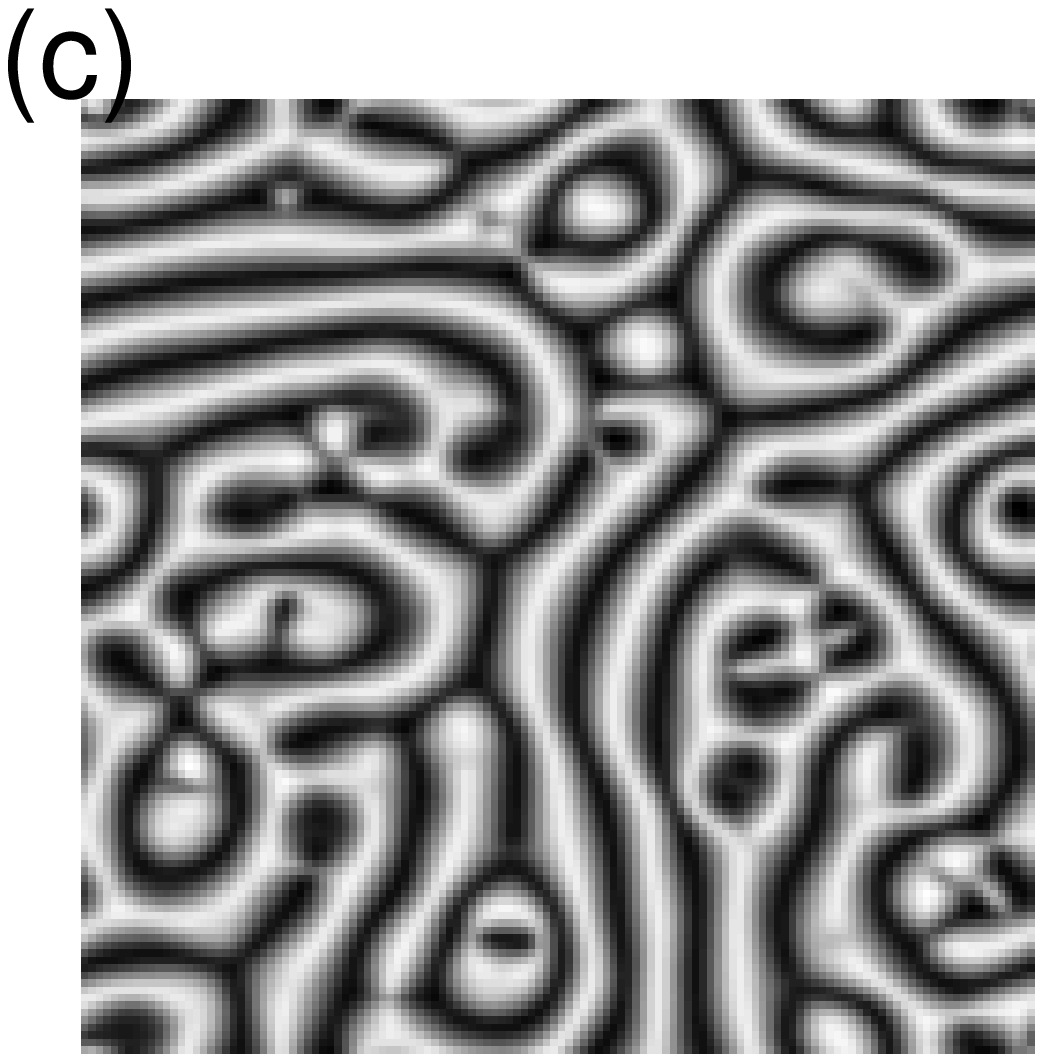}}
{\includegraphics[width=4.2cm]{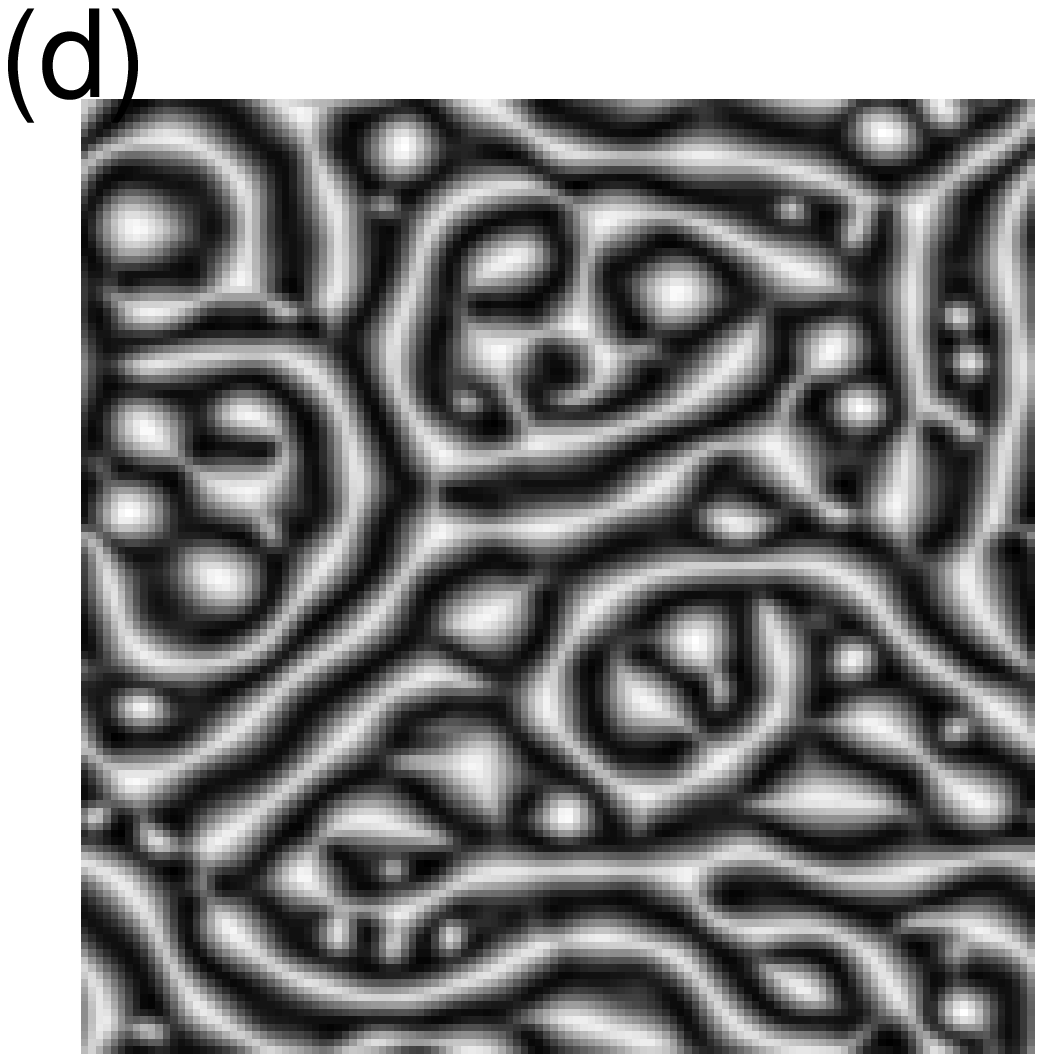}}
\end{center}%
\end{figure}

{\bf Figure 1.}

\vfill\eject

\begin{figure}%
\begin{center}%
{\includegraphics[width=4.2cm]{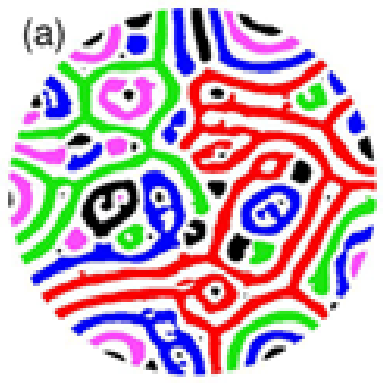}}
{\includegraphics[width=4.2cm]{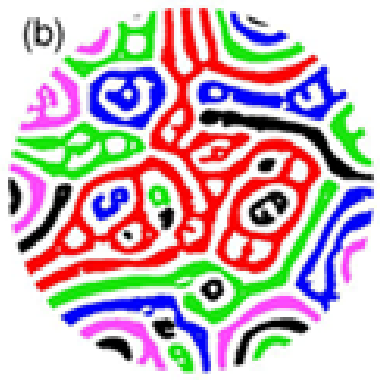}}\\
{\includegraphics[width=4.2cm]{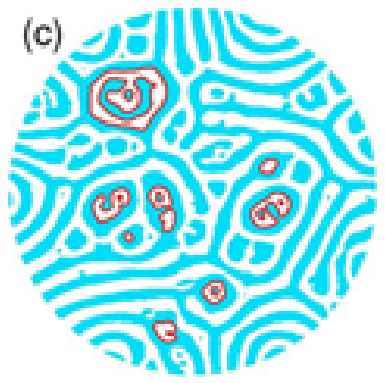}}
{\includegraphics[width=4.2cm]{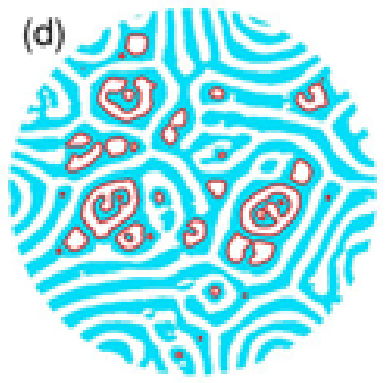}}
\end{center}%
\end{figure}

{\bf Figure 2.}

\vfill\eject

\begin{figure}%
\begin{center}%
{\includegraphics[width=6.5cm,height=3.5cm]{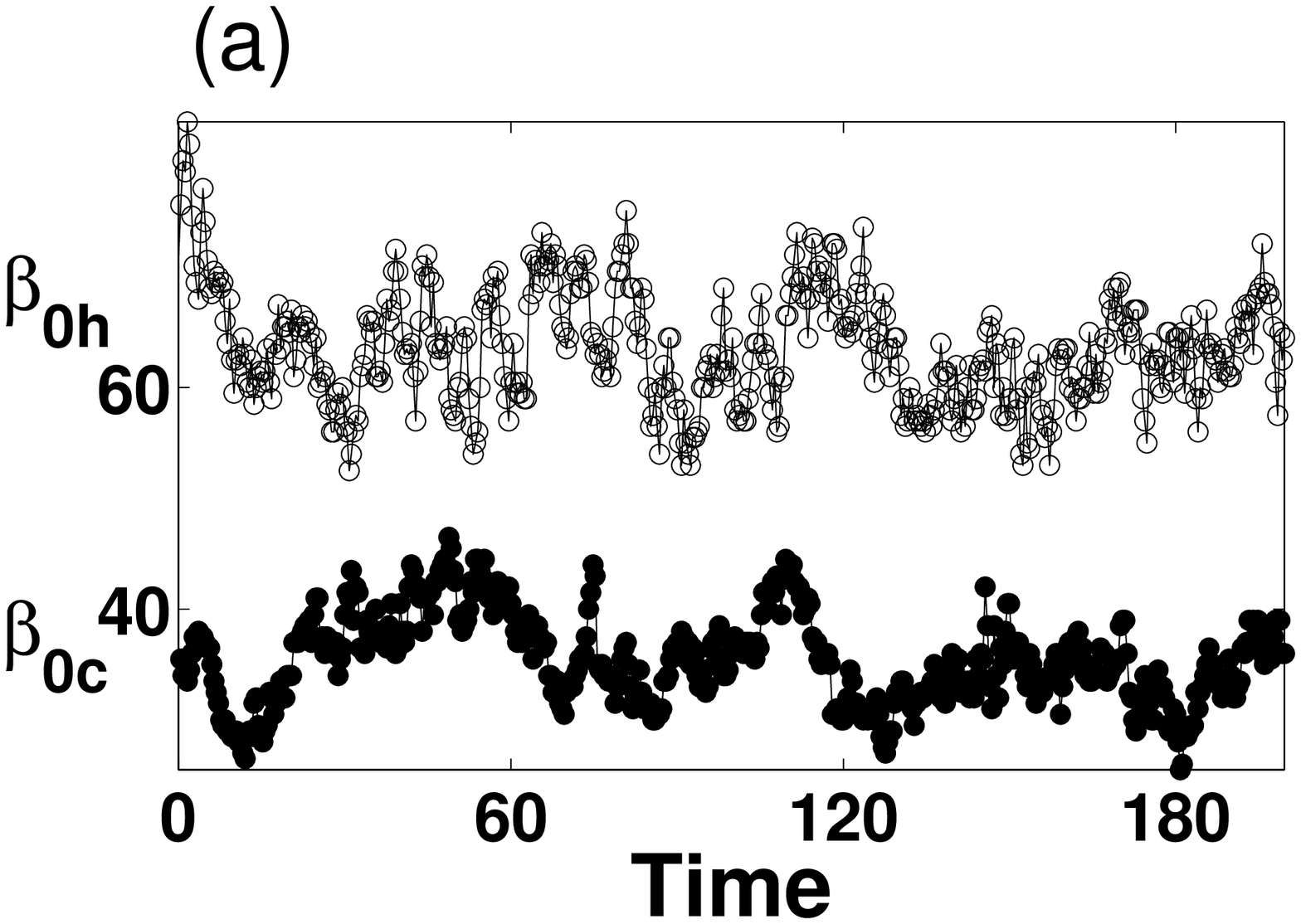}}
{\includegraphics[width=6.5cm,height=3.5cm]{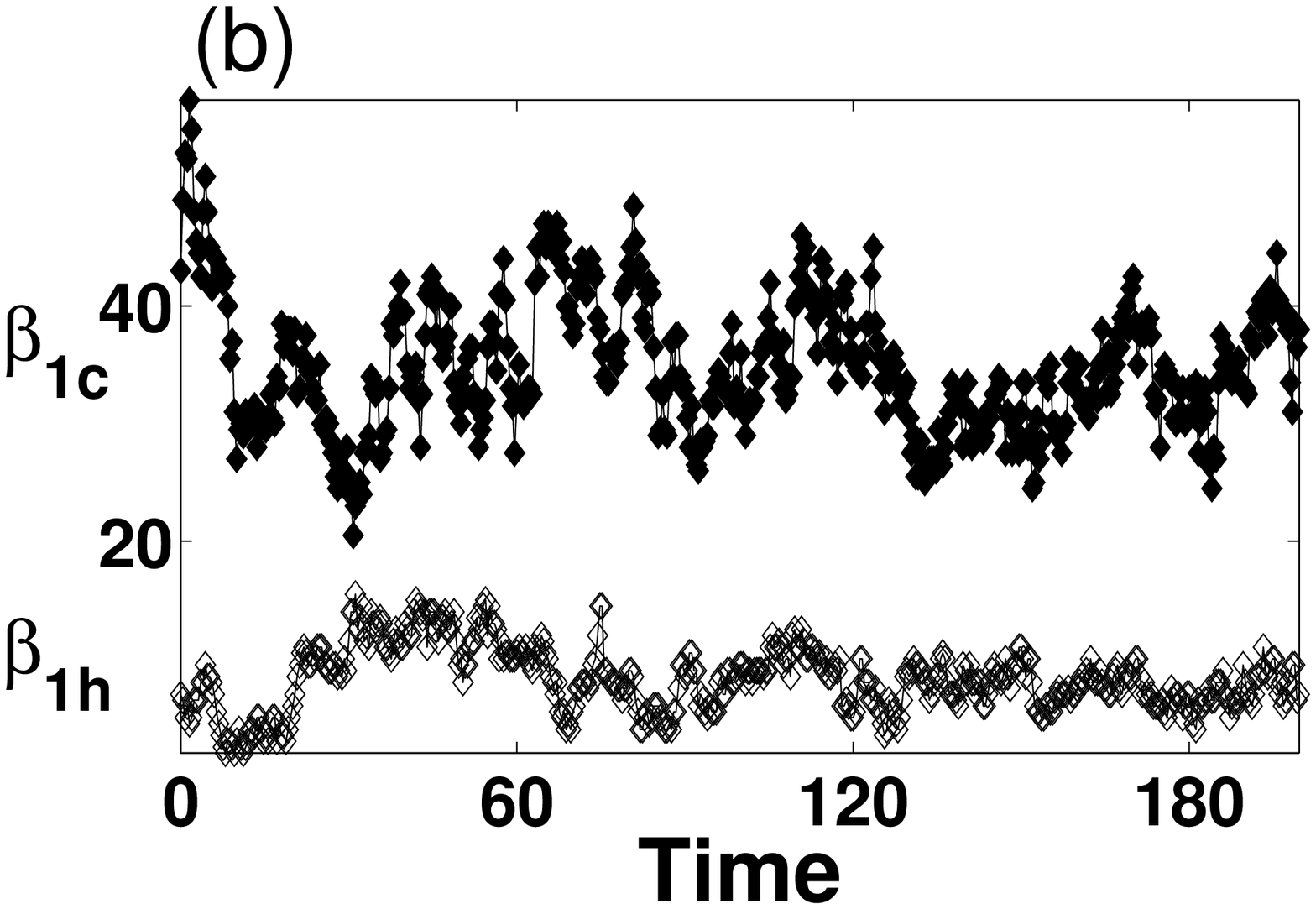}}
\end{center}%
\end{figure}

{\bf Figure 3.}

\vfill\eject

\begin{figure}%
\begin{center}%
{\includegraphics[width=8cm]{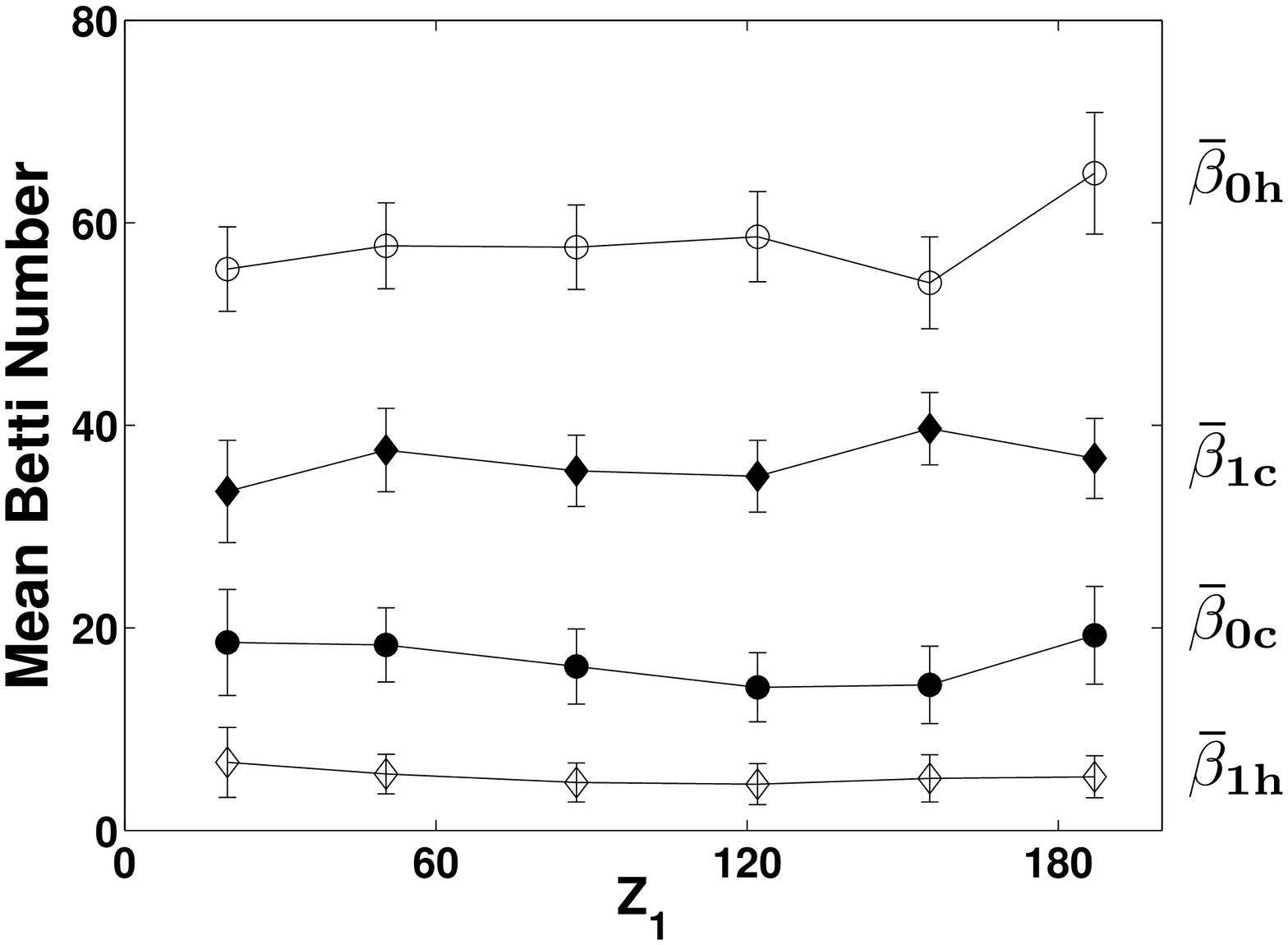}}
\end{center}%
\end{figure}

{\bf Figure 4.}

\vfill\eject

\begin{figure}%
\begin{center}%
{\includegraphics[width=8cm]{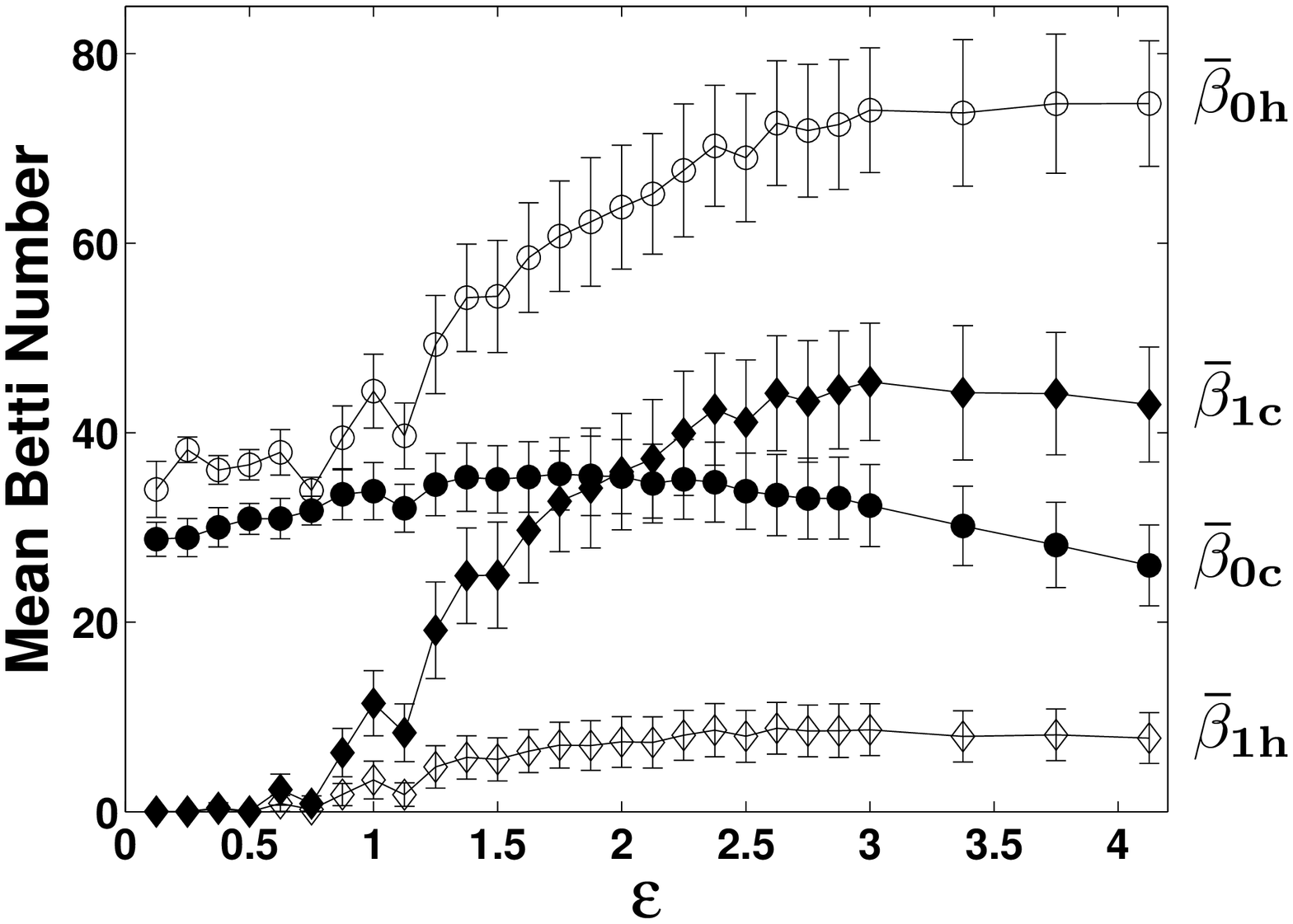}}
\end{center}%
\end{figure}

{\bf Figure 5.}

\vfill\eject

\begin{figure}%
\begin{center}%
{\includegraphics[width=6.5cm,height=3.5cm]{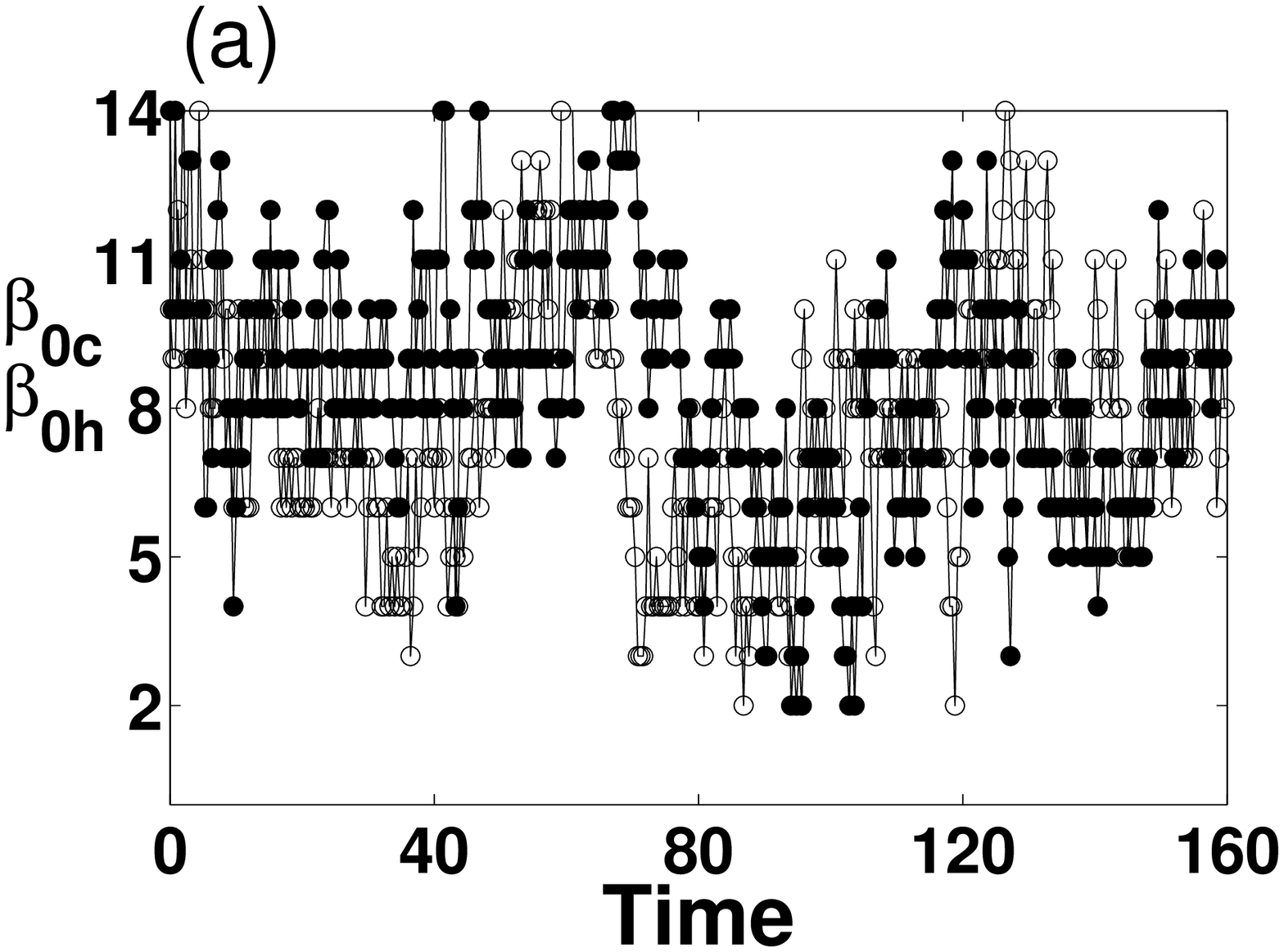}}
{\includegraphics[width=6.5cm,height=3.5cm]{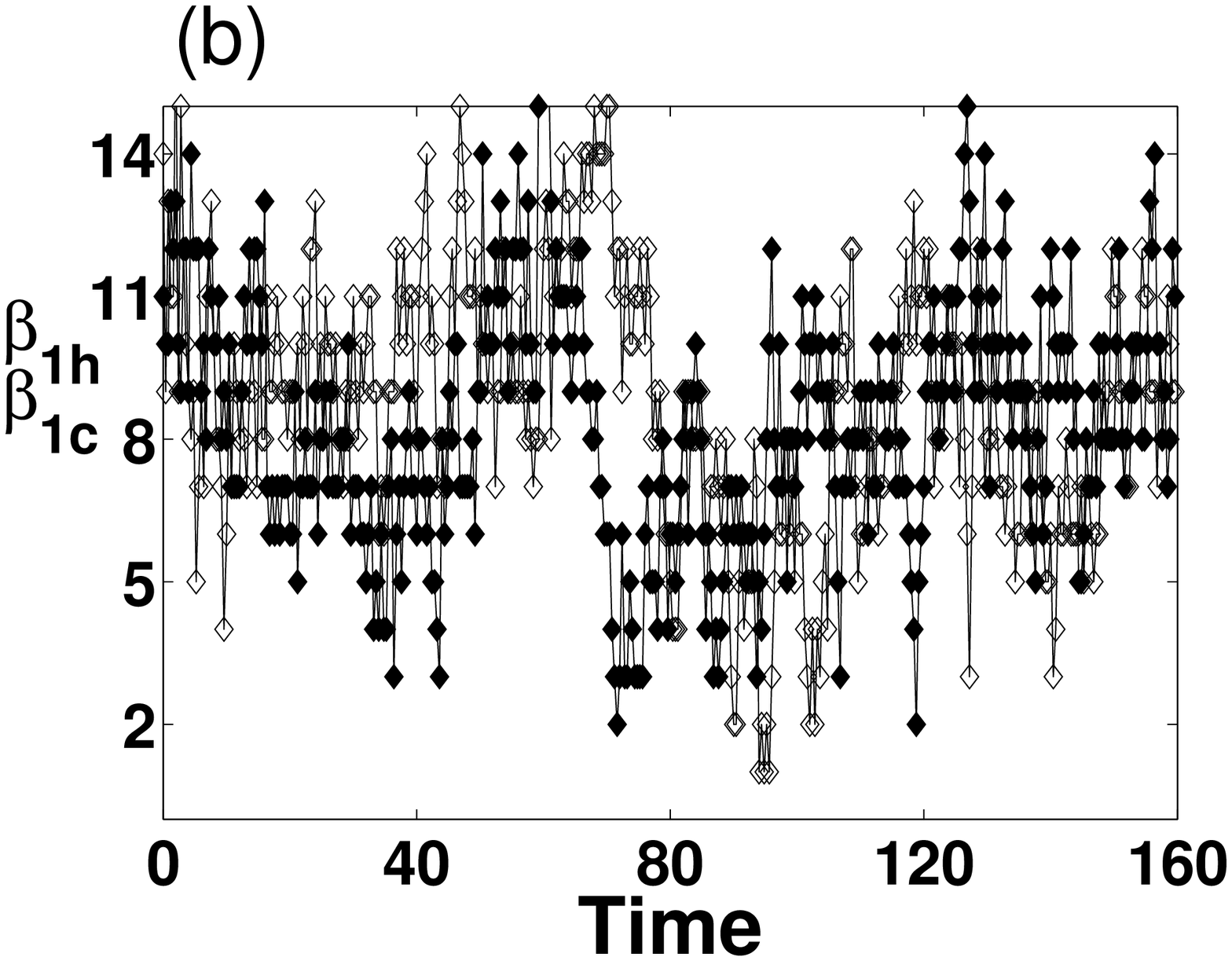}} \\
{\includegraphics[width=6.5cm,height=3.5cm]{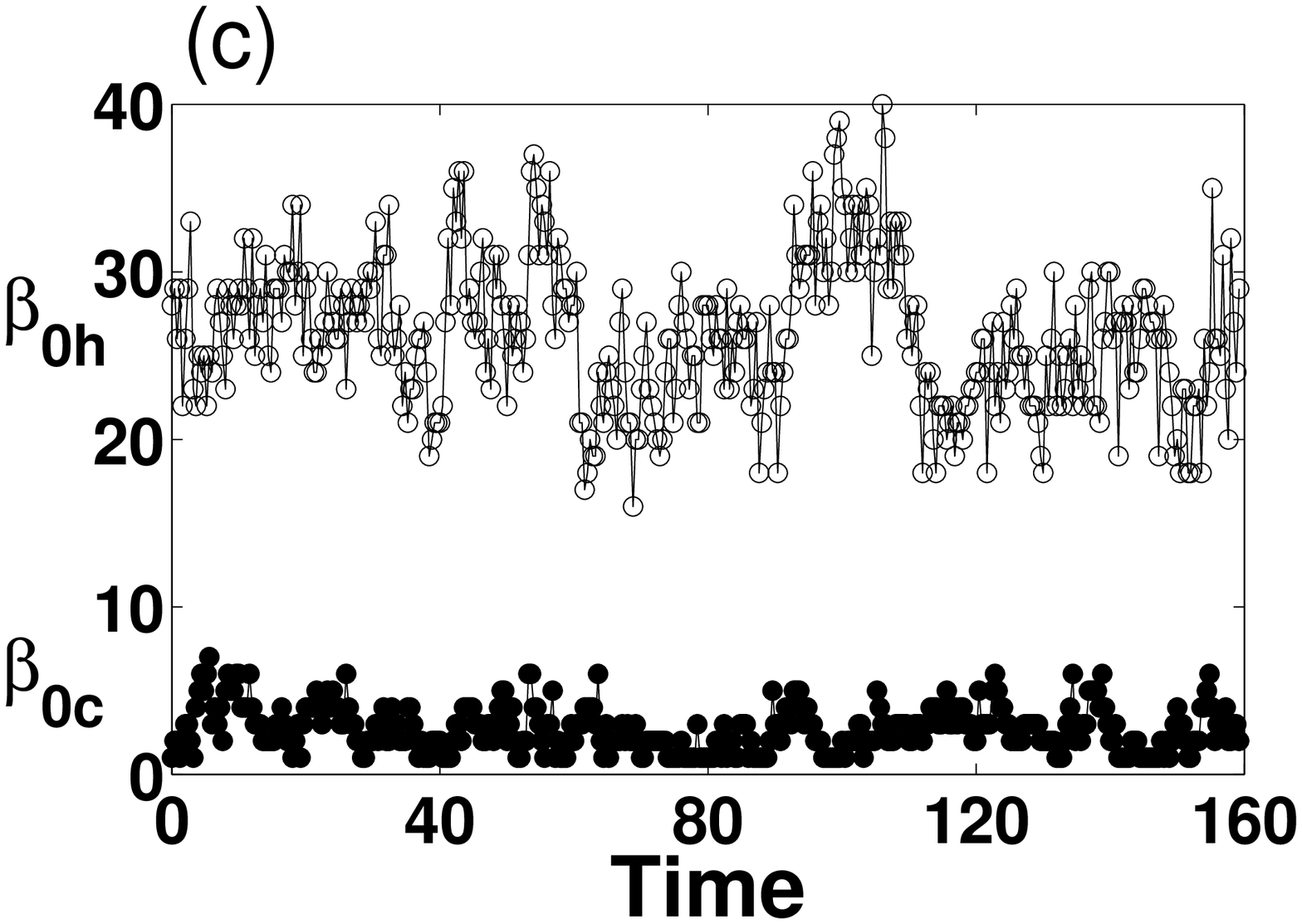}}
{\includegraphics[width=6.5cm,height=3.5cm]{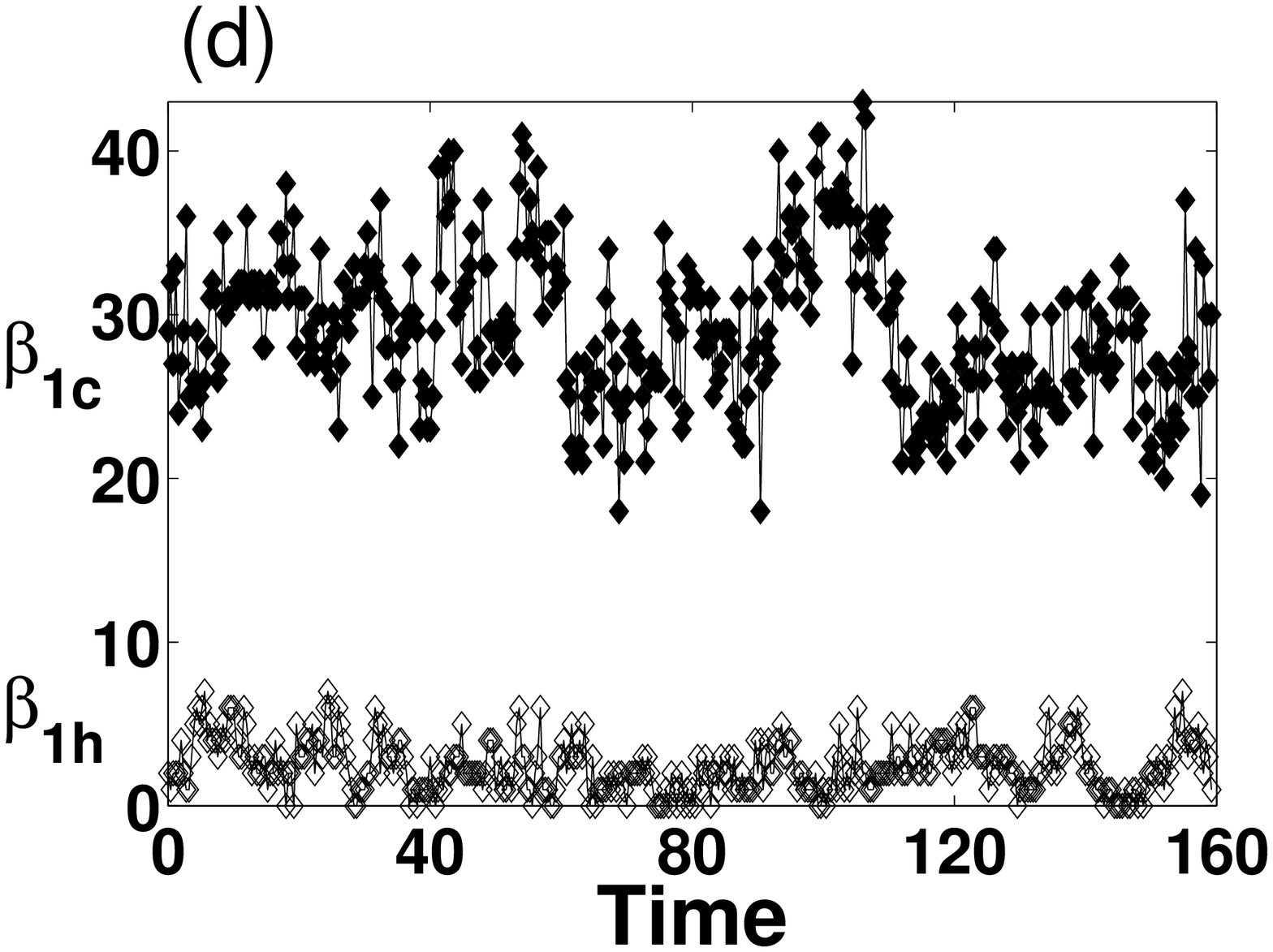}} \\
{\includegraphics[width=6.5cm,height=3.5cm]{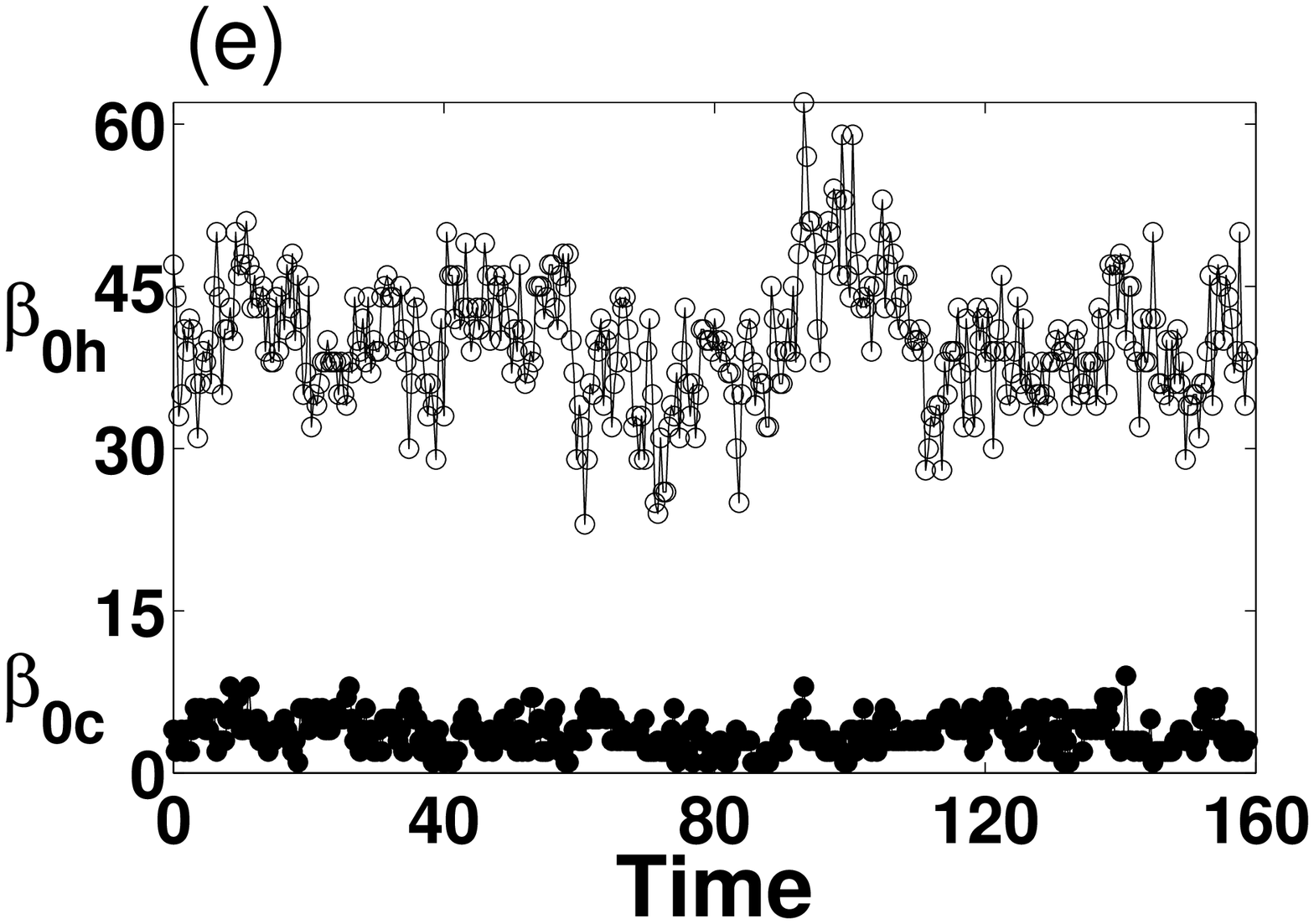}}
{\includegraphics[width=6.5cm,height=3.5cm]{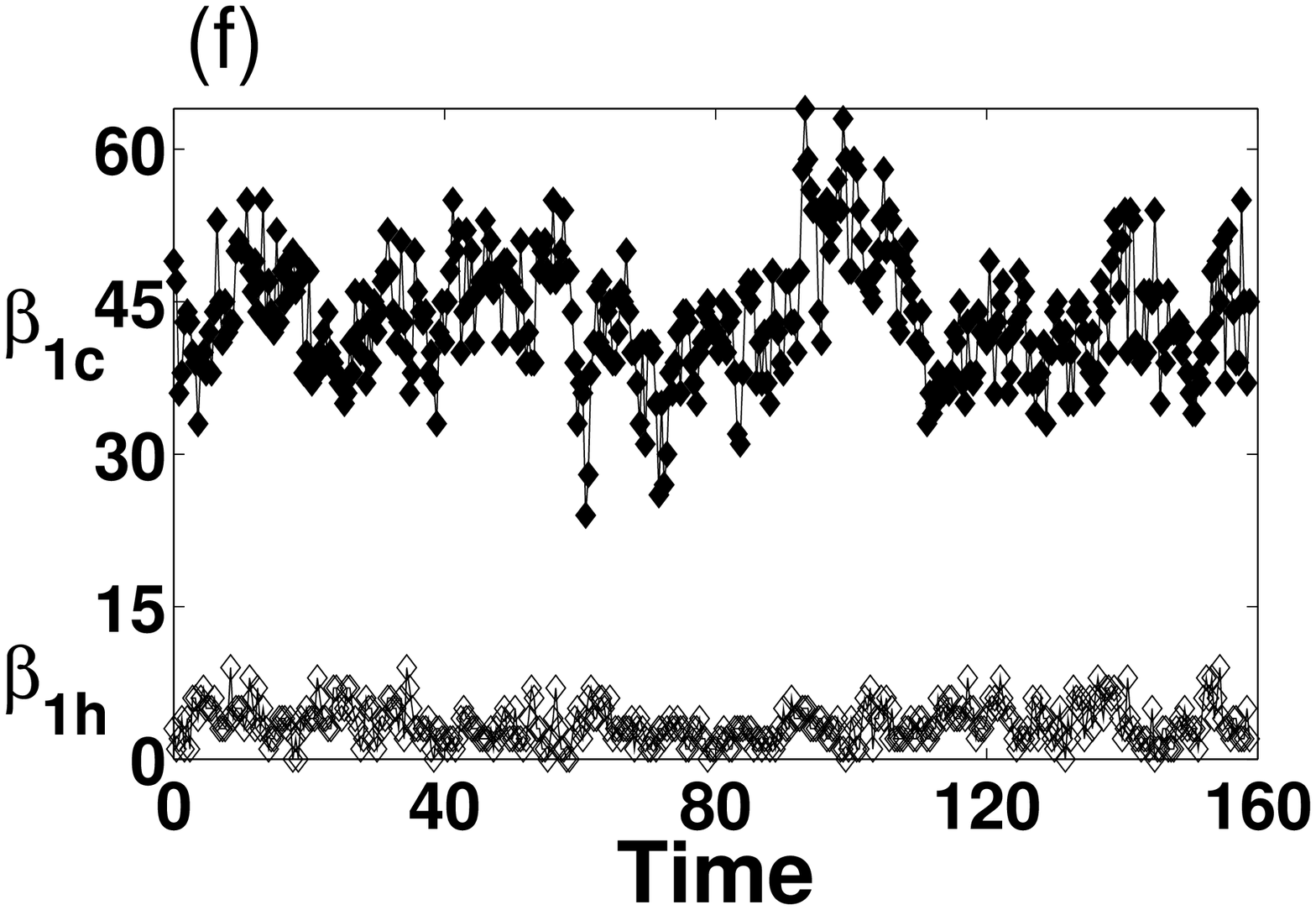}}
\end{center}%
\end{figure}

{\bf Figure 6.}

\end{document}